\documentclass[twocolumn,a4paper]{sn-jnl}

\usepackage{epsfig}
\usepackage{hhline}
\usepackage{amsmath}
\usepackage{amssymb}
\usepackage{mathrsfs}
\usepackage{tabularx}
\usepackage{placeins}

\usepackage{subcaption}
\newcolumntype{s}{>{\hsize=.5\hsize}X}

\usepackage{ bbold }

\usepackage{color}
\usepackage{xspace}
\usepackage[percent]{overpic}
\usepackage{paralist} 
\usepackage{caption}
\usepackage{bm} 
\usepackage{acronym}

\usepackage{hyperref} 
\hypersetup{colorlinks=true, urlcolor=blue}
\usepackage{cite} 
\hypersetup{
  citecolor=[rgb]{0.,0.,0.5},
  linkcolor=[rgb]{0.6,0.,0.0},
  urlcolor=[rgb]{0.,0.,0.5}
  }
\usepackage{cleveref}
\usepackage{booktabs} 
\usepackage{xcolor}

\usepackage{parskip}

\usepackage[utf8]{inputenc}
\newlength{\dinwidth}
\newlength{\dinmargin}
\begin{document}
\pagestyle{empty}

\newcommand{\GeVsq}{\ensuremath{\mathrm{GeV}^2}}
\newcommand{\GeV}{\ensuremath{\mathrm{GeV}} }
\newcommand{\pt}{\ensuremath{p_{T}}}
\newcommand{\PP}{\ensuremath{\mathcal{P}}}
\newcommand{\Qsq}{\ensuremath{Q^{2}}\xspace}
\newcommand{\etaB}{\ensuremath{\eta_{\mathrm{_{Breit}}}}\xspace}
\newcommand{\etaL}{\ensuremath{\eta_{\mathrm{_{Lab}}}}\xspace}
\newcommand{\xB}{\ensuremath{x_\mathrm{Bj}\xspace}}
\newcommand{\chisq}{\ensuremath{\chi^{2}}}
\newcommand{\hchisq}{\ensuremath{\hat\chi^{2}}}
\newcommand{\chisqA}{\ensuremath{\chi_{\rm A}^{2}}}
\newcommand{\chisqL}{\ensuremath{\chi_{\rm L}^{2}}}
\newcommand{\ndf}{\ensuremath{n_{\rm dof}}}

\newcommand{\V}{\ensuremath{\mathbb{V}}}
\newcommand{\W}{\ensuremath{\mathbb{W}}}
\newcommand{\M}{\ensuremath{\tilde{M}}}
\newcommand{\Y}{\ensuremath{Y\M}}
\newcommand{\U}{\ensuremath{\Upsilon}}

\newcommand{\dt}{\ensuremath{\bm{d}}}
\newcommand{\mb}{\ensuremath{\bm{\bar{m}}}}
\newcommand{\mt}{\ensuremath{\bm{\tilde{m}}}}
\newcommand{\s}{\ensuremath{\bm{s}}}
\newcommand{\ahut}{\ensuremath{\bm{\hat{a}}}}

\newcommand{\ptjet}{\ensuremath{p_{\rm T}^{\rm jet}}}
\newcommand{\Mjj}{\ensuremath{m_{12}}}
\newcommand{\mz}{\ensuremath{m_{\rm Z}}\xspace}
\newcommand{\as}{\ensuremath{\alpha_{\rm s}}\xspace}
\newcommand{\asmz}{\ensuremath{\as(\mz)}\xspace}
\newcommand{\asmzPDF}{\ensuremath{\as^{\rm PDF}(\mz)}\xspace}
\newcommand{\asmzf}{\ensuremath{\as^{\Gamma}(\mz)}\xspace}
\newcommand{\asmur}{\ensuremath{\alpha_{\rm s}(\mur)}\xspace}
\newcommand{\chad}{\ensuremath{c_{\rm had}}\xspace}
\newcommand{\ord}{\ensuremath{\mathcal{O}}\xspace}
\newcommand{\PDFasResult}{0.1142}
\newcommand{\HonePDF}{H1PDF2017\,{\protect\scalebox{0.8}{[NNLO]}}}
\newcommand{\TODO}{\color{red} Todo.}
\newcommand{\NEW}{\color{blue}}

\newcommand{\tQ}{\ensuremath{\tau_Q}\xspace}
\newcommand{\tb}{\ensuremath{\tau^{b}_{1_{\text{Gr.}}}}\xspace}
\newcommand{\GIM}{\ensuremath{M_{\text{Gr.}}^{2}/Q^2}\xspace}
\newcommand{\pzb}{\ensuremath{P_{z}^\text{Breit}}\xspace}
\newcommand{\pzbi}{\ensuremath{P_{z,i}^\text{Breit}}\xspace}


\definecolor{myOrange}{rgb}{1,0.5,0.}
\definecolor{myGreen}{rgb}{0.0,0.6,0.1}
\newcommand{\redtext}[1]{\textcolor{red}{#1}}
\newcommand{\bluetext}[1]{\textcolor{blue}{#1}}
\newcommand{\greentext}[1]{\textcolor{myGreen}{#1}}
\newcommand{\orangetext}[1]{\textcolor{myOrange}{#1}}

\newcommand{\commenting}[2]{\redtext{(#1: #2)}}
\newcommand{\reply}[2]{\greentext{(#1: #2)}}
\newcommand{\followup}[2]{\bluetext{(#1: #2)}}
\newcommand{\suggestion}[2]{\orangetext{(#1: #2)}}

\newcommand{\sqrts}{\ensuremath{\sqrt{s}}\xspace}
\newcommand{\emp}{\ensuremath{e^{-}p}\xspace}
\newcommand{\epp}{\ensuremath{e^{+}p}\xspace}
\newcommand{\ep}{\ensuremath{ep}\xspace}
\newcommand{\epem}{\ensuremath{e^{+}e^{-}}\xspace}

\newcommand{\invpb}{\ensuremath{\mathrm{pb}^{-1}}\xspace}
\newcommand{\alphas}{\ensuremath{\alpha_{S}}\xspace}
\newcommand{\pp}{\ensuremath{pp\xspace}}

\newcommand{\Lint}{\ensuremath{\mathcal{L}_\mathrm{int}\xspace}}

\newcommand{\pT}{\ensuremath{p_\mathrm{T}}\xspace}

\newcommand{\kT}{\ensuremath{k_\mathrm{T}}\xspace}

\newcommand{\zg}{\ensuremath{z_{\mathrm{g}}\xspace}}
\newcommand{\zcut}{\ensuremath{z_{\mathrm{cut}}\xspace}}
\newcommand{\zi}{\ensuremath{z_{i}\xspace}}
\newcommand{\zj}{\ensuremath{z_{j}\xspace}}

\newcommand{\antikT}{\ensuremath{\mathrm{anti}-k_\mathrm{T}\xspace}}

\newcommand{\Mgrsq}{\ensuremath{M_{\text{Gr.}}^{2}}}
\newcommand{\Qminsq}{\ensuremath{Q^2_\mathrm{min.}}}

\newcommand{\GIMa}{\ensuremath{M_{\text{Gr.}}^{2}/\Qminsq}\xspace}
\newcommand{\GIMln}{\ensuremath{\mathrm{ln}(M_{\text{Gr.}}^{2}/\Qminsq)}\xspace}
\newcommand{\GIMlnone}{\ensuremath{\mathrm{ln}(M_{\text{Gr.}}^{2}/150~\mathrm{GeV}^2)}\xspace}

\newcommand{\tauoneb}{\ensuremath{\tau^{b}_{1}}}
\newcommand{\tauonebgr}{\ensuremath{\tau^{b}_{1,{\text{Gr.}}}}}

\received{March 2024}
\printhistory
\motto{}{Published in EPJC \hfill DESY-24-036}

\title{Measurement of groomed event shape observables in
  deep-inelastic electron-proton scattering at HERA
}
\subtitle{H1 Collaboration}

\author[50]{\fnm{V.}\sur{Andreev}}
\author[34]{\fnm{M.}\sur{Arratia}}
\author[46]{\fnm{A.}\sur{Baghdasaryan}}
\author[9]{\fnm{A.}\sur{Baty}}
\author[40]{\fnm{K.}\sur{Begzsuren}}
\author[17]{\fnm{A.}\sur{Bolz}}
\author[30]{\fnm{V.}\sur{Boudry}}
\author[15]{\fnm{G.}\sur{Brandt}}
\author[26]{\fnm{D.}\sur{Britzger}}
\author[7]{\fnm{A.}\sur{Buniatyan}}
% ...
\affil[7]{\orgaddress{School of Physics and Astronomy, University of Birmingham, Birmingham, United Kingdom$^{f3}$}}
% ...
\affil[9]{\orgaddress{University of Illinois, Chicago, IL 60607, USA}}
% ...
\affil[15]{\orgaddress{II. Physikalisches Institut, Universität Göttingen, Göttingen, Germany}}
% ...
\affil[17]{\orgaddress{Deutsches Elektronen-Synchrotron DESY, Hamburg and Zeuthen, Germany}}
% ...
\affil[26]{\orgaddress{Max-Planck-Institut für Physik, München, Germany}}
% ...
\affil[30]{\orgaddress{LLR, Ecole Polytechnique, CNRS/IN2P3, Palaiseau, France}}
% ...
\affil[34]{\orgaddress{University of California, Riverside, CA 92521, USA}}
% ...
\affil[40]{\orgaddress{Institute of Physics and Technology of the Mongolian Academy of Sciences, Ulaanbaatar, Mongolia}}
% ...
\affil[46]{\orgaddress{Yerevan Physics Institute, Yerevan, Armenia}}
% ...
\affil[50]{\orgaddress{Affiliated with an institute covered by a current or former collaboration agreement with DESY}}

\abstract{\unboldmath%
The H1 Collaboration at HERA reports the first measurement of groomed event shape observables in deep inelastic electron-proton scattering (DIS) at $\sqrts=319~$GeV, using data recorded between the years 2003 and 2007 with an integrated luminosity of 351 \invpb. Event shapes provide incisive probes of perturbative and non-perturbative QCD. Grooming techniques have been used for jet measurements in hadronic collisions; this paper presents the first application of grooming to DIS data. The analysis is carried out in the Breit frame, utilizing the novel Centauro jet clustering algorithm that is designed for DIS event topologies. Events are required to have squared momentum-transfer $Q^2 > 150$ GeV$^2$ and inelasticity $ 0.2 < y < 0.7$. We report measurements of the production cross section of groomed event 1-jettiness and groomed invariant mass for several choices of grooming parameter. Monte Carlo model calculations and analytic calculations based on Soft Collinear Effective Theory are compared to the measurements.
}

\maketitle


\section{Introduction}
\label{sect:intro}

Event shape observables characterize the distribution of final-state particles produced in high-energy particle interactions. Event shapes have been measured extensively in \epem\ collisions~\cite{Hanson:1975fe,TASSO:1979zyf,JADE:1981ofk,DELCO:1985qwh,CELLO:1987vag,Bender:1984fp,DELPHI:1996sen,OPAL:1997asf,ALEPH:1996oqp,OPAL:1990xiz,L3:1992nwf,OPAL:2004wof} and in \ep\ collisions~\cite{Adloff:1997gq,Adloff:1999gn,Aktas:2005tz,ZEUS:2002tyf,ZEUS:2006vwm}; such observables are calculable to high precision using perturbative Quantum Chromodynamics (pQCD)~\cite{Knobbe:2023ehi,Dokshitzer:1995zt,Dasgupta:1997ex,Antonelli:1999kx,Dasgupta:2001sh}. Event shapes are incisive probes of QCD, notably to constrain the strong coupling constant $\alpha_{s}$~\cite{ALEPH:1996oqp,SLD:1994idb,Becher:2008cf,Gehrmann-DeRidder:2007vsv,Dissertori:2009ik,Bethke:2009jm,Hoang:2015hka,Marzani:2019evv}. The description of hadronic final states in Monte Carlo event generators has likewise benefitted substantially from event shape measurements~\cite{DELPHI:1996sen,Knobbe:2023ehi,Skands:2010ak,Ilten:2016csi,Skands:2014pea,LaCagnina:2023yvi}. 

Jets arise from energetic partons (quarks and gluons) produced in hard interactions. The partons are initially highly virtual, decaying in a partonic shower that is experimentally observable as a correlated spray of hadrons. Jets provide a laboratory for testing QCD~\cite{Salam:2010nqg}. However, the precision of jet measurements at hadron colliders is limited by the contribution of non-perturbative (NP) processes and by the presence of the underlying event, which consists of final-state particles that do not originate from the hard-scattering process that produced the jet being studied. This limitation is addressed by the application of jet grooming algorithms~\cite{Butterworth:2008iy,Thaler:2010tr,Dasgupta:2013ihk,Larkoski:2014wba,Kang:2018vgn,Larkoski:2017jix,Kogler:2021kkw,H1:2023fzk}, which systematically removes particles likely to originate in NP processes and the underlying event, in a way that is theoretically and experimentally well-controlled~\cite{Kogler:2018hem}. 

In jet grooming algorithms, typically the Cambridge-Aachen~\cite{CMS:2009lxa} sequential recombination algorithm is applied, which combines jet constituents (particle four-vectors) with small angular distance into jets in the first recombination steps, and those with wide angular distance in later steps. The recombination sequence is then inspected in reverse, starting with the softest branch. For each step in this inspection the kinematics of its contributing branches are compared to a specified condition; for instance, the ratio \zg\ of the transverse momentum (\pT) of the softer branch to the summed \pT\ of both branches is required to satisfy $\zg>\zcut$, for a given value of \zcut. If this condition is not satisfied, the softer branch is discarded from the event (hence the term ``grooming'') and the comparison continues to the next recombination step along the harder branch until the condition is satisfied. If no recombination step satisfies the condition, the event is discarded. The choice of condition differentiates jet grooming algorithms~\cite{Kogler:2018hem}.


 The technique of jet grooming has been applied extensively in proton-proton (\pp) collisions at the Large Hadron Collider and at the Relativistic Heavy Ion Collider, for instance to search for the decay of boosted heavy particles, discriminate jets initiated by quarks or gluons, or measure jet substructure~\cite{ALICE:2022hyz,ATLAS:2013bqs,ATLAS:2014bjq,ATLAS:2015ull,CMS:2018fof,CMS:2018vzn,ALICE:2019ykw,ATLAS:2019dty,ATLAS:2019mgf,STAR:2021lvw,ALICE:2021vrw,ALICE:2021njq,Cunqueiro:2022svx,ALICE:2022phr,Kogler:2018hem}. Jet grooming has also been used to study the modification of jets propagating in the Quark-Gluon Plasma generated in heavy-ion collisions~\cite{CMS:2017qlm,CMS:2018fof,ALargeIonColliderExperiment:2021mqf,STAR:2021kjt, ATLAS:2022vii,ALICE:2023dwg}.

Jets have also been measured extensively in \epem~\cite{JADE:1986kta,TASSO:1983cre,OPAL:1989fep,DELPHI:1990svh,DELPHI:2005wpx,OPAL:1995cgp,ALEPH:1996oqp} and lepton-proton deep inelastic scattering (DIS)~\cite{PhysRevLett.69.1026,PhysRevLett.72.466,Adloff:1998gg,H1:2000bqr,H1:1998rpm,H1:1995tux,H1:1998cuj,H1:2009pqp,H1:2007xjj,H1:2010mgp,H1:1994lps,H1:2002qhb,ZEUS:2005iex,ZEUS:2002nms,ZEUS:1994jfw,ZEUS:1995tgg,ZEUS:2006xvn,H1:2014cbm,Collaboration:2023xha}. The underlying event background in such collision systems is smaller than in \pp\ collisions,
enabling more accurate measurement of soft components of jets. In addition, the kinematic properties of the partonic scattering are known experimentally, providing more precise comparison to QCD calculations. To date, grooming has not been applied to data from \ep\ DIS. 

It was recently proposed to apply grooming algorithms not only to jet measurements at hadron colliders, but also to full events in $ep$ DIS~\cite{Makris:2021drz}, leveraging the similarities between DIS at high virtualities and single jets.
From a theoretical standpoint, groomed observables in \ep\ DIS are free of non-global logarithms~\cite{Makris:2021drz,Dasgupta:2001sh,Marzani:2017mva,Larkoski:2015npa} and have reduced hadronization effects relative to ungroomed observables~\cite{Larkoski:2014wba,Marzani:2017mva}. In addition, the magnitude of the non-perturbative component of such observables is controllable experimentally by varying the strength of the grooming parameter \zcut~\cite{Larkoski:2014wba,Makris:2021drz,Marzani:2017mva}.


In this paper, the H1 Collaboration at HERA reports the first measurement of groomed event shapes in \emp\ and \epp\ neutral-current (NC) deep inelastic scattering (electrons and positrons are referred to generically as ``electrons'' in the following). The data were recorded during the years 2003 to 2007 with electron and proton beam energies of $E_e=27.6$ GeV and $E_p=920$ GeV, respectively, corresponding to $\sqrts=319$ GeV. The recorded dataset has integrated luminosity $\Lint=351~\invpb$. The analysis is based on events with exchanged-boson virtuality $Q^2 > 150$ GeV$^2$ and inelasticity $ 0.2 < y < 0.7$. We report measurements of the production cross section of groomed event 1-jettiness and groomed event invariant mass for several choices of grooming parameter \zcut. Monte Carlo model calculations and analytic calculations based on Soft Collinear Effective Theory (SCET)~\cite{Makris:2021drz} and NNLO+NLL$^{\prime}$ pQCD~\cite{Knobbe:2023ehi} are compared to the measurements.

These data provide new, differential constraints on the detailed structure of DIS-induced final states, which are valuable for the tuning of MC event generators. Improvement of such event generators is important, for instance, for the physics program of the future Electron-Ion Collider~\cite{AbdulKhalek:2021gbh}.


The paper is organized as follows:
Section~\ref{sect:ExpDataset} describes the experimental apparatus and dataset;
Section~\ref{sect:Theory} presents theoretical calculations that are compared to the data;
Section~\ref{sect:Analysis} outlines the analysis procedure and event selection;
Section~\ref{sect:Corrections} presents the corrections applied to the data;
Section~\ref{sect:Uncertainties} presents the uncertainties;
Section~\ref{sect:Results} reports results of the analysis and the comparison of theoretical predictions to the measurements; and 
Section~\ref{sect:Summary} provides a summary.

\section{Experimental Setup}
\label{sect:ExpDataset}

The H1 experiment is a general purpose particle detector with full azimuthal coverage around the electron-proton interaction region~\cite{H1:1996prr,H1CalorimeterGroup:1993boq,H1:1996prr,H1:1996jzy,H1SPACALGroup:1996ziw,Pitzl:2000wz}. The detector is described in Refs.~\cite{Abt:1996hi,Abt:1996xv,Pitzl:2000wz,Appuhn:1996na}. H1 employs a right-handed coordinate system in which the proton beam direction defines the positive $z$ direction. The nominal $ep$ interaction point is located at $z=0$. 

The liquid argon calorimeter (LAr)
\cite{H1CalorimeterGroup:1993uzc,H1CalorimeterGroup:1994gfk}, which provides a trigger for high-$Q^2$ neutral current DIS, subtends polar angles of $4^\circ <\theta < 154^\circ$ and provides an energy resolution of $\sigma(E) /E \simeq (11\%/\sqrt{E/\GeV}) \oplus 1\%$ for electrons and $\sigma(E)/E \simeq (55\%/ \sqrt{E/\GeV}) \oplus 3\%$ for charged pions. The central tracking system, comprising gaseous drift and proportional chambers and a silicon vertex detector, covers the polar range $15^\circ<\theta<165^\circ$ and has transverse momentum resolution for charged particles $\sigma(\pT)/{\pT}\simeq 0.2\% \cdot \pT/\GeV \oplus 1.5\%$. A lead-scintillating fiber calorimeter (SpaCal)~\cite{H1CalorimeterGroup:1993boq,H1SPACALGroup:1995xvh}, consisting of both electromagnetic and hadronic sections, covers the backward direction ($153^\circ < \theta < 177^\circ$). The SpaCal has electromagnetic energy resolution of $\sigma(E)/{E} \simeq (7.5\%/\sqrt{E/\GeV})\oplus 2\%$. 

Events are triggered by requiring a high-energy cluster in the electromagnetic portion of the LAr. The trigger and event selection used in the analysis closely follows Ref.~\cite{H1:2014cbm}. The efficiency of the trigger is greater than $99\%$ in the phase space of this analysis. Events are further selected online by requiring the scattered electron to have an energy greater than 11 GeV and to fall within the LAr fiducial volume. 

In the offline analysis, an energy flow algorithm combines information from the tracking detectors and calorimeters to generate a set of four-vectors~\cite{energyflowthesis,energyflowthesis2,energyflowthesis3}. These four-vectors are used to define the scattered electron and the hadronic final state (HFS). Isolated energy deposits with high energy in the backward and central sections of the electromagnetic calorimeters are typically the result of QED radiation of a photon off the electron. If the energy deposit is closer to the scattered electron than the electron beam direction, the photon is likely to come from final-state radiation and its four-vector is recombined with the four-vector of the scattered electron. If the photon is closer to the electron beam ($-z$) direction, it is likely the result of initial-state radiation (ISR) and is removed from the event. The HFS is defined as all the particles remaining after this procedure, excluding the scattered electron.

The kinematic variables describing the event, $Q^2$, $y$, and Bjorken $x$ (denoted \xB), are reconstructed using the $I\Sigma$ method as described in Ref. \cite{Bassler:1994uq}. Bjorken $\xB=\Qsq/(2P\cdot q)$, where $P$ and $q$ refer to the incoming proton and exchanged boson four-vectors, respectively. The inelasticity $y=P\cdot q / P\cdot k$, where $k$ is the incoming electron four-vector. These kinematic variables are used to reconstruct the boost to the Breit frame, as well as the vector defining the current hemisphere of the Breit frame, $q_{J}=xP+q$. The definitions of the Breit frame and $q_J$ are described in more detail in Section~\ref{sect:Analysis}. 

Events are further selected based on the following quality assurance cuts placed on quantities measured in the laboratory frame:
\begin{compactitem}
\item The measured $z$-location of the event vertex is constrained to be within 35 cm of the nominal vertex $z$-location. 
\item The total longitudinal momentum balance ($E-p_z$) of the event, evaluated by summing the $E-p_z$ of all measured particles, is required to satisfy $50 < E-p_z < 60$ GeV. This requirement predominantly serves to reduce the contribution from events with significant QED initial-state radiation and events likely to come from photoproduction or beam-gas background. It has the additional benefit of rejecting events in which the hadronic final state is poorly reconstructed.
\item The total transverse momentum (\pT) of the hadronic final state is required to approximately balance that of the scattered electron \pT, $0.6 < {p_\mathrm{T,HFS}}/{p_{\mathrm{T},e}} < 1.6$ and ${p_\mathrm{T,HFS}}-{p_{\mathrm{T},e}} < 5$ GeV. These requirements ensure that the hadronic final state and scattered electron are well-measured.
\item The vector defining the current hemisphere of the Breit frame, $q_{J}$, is required to have a polar angle $7^{\circ}<\theta_{q_J}<175^{\circ}$ in the lab frame in order to suppress non-collision backgrounds and to ensure the hadronic final state is properly contained within the detector. 
\item Events with $\theta_{q_J}>149^{\circ}$ and the velocity of the Lorentz boost to the Breit frame $\beta> 0.9$ are rejected to suppress contamination from initial-state QED radiation. Events in which the difference between the polar angles of the HFS and the current hemisphere vector satisfies $\theta_{q_J} - \theta_{\mathrm{HFS}} > 90^{\circ}$ are also rejected for the same reason.
\item In the kinematic region studied here, the hadronic final state and the scattered electron are typically produced at similar polar angles. Events events with $Q^2 < 700$ GeV$^2$ and $|\Delta\eta|>0.3$ are poorly measured and are rejected, where pseudorapidity in the laboratory frame $\eta=-\mathrm{ln}(\mathrm{tan}\frac{\theta}{2})$, and $\Delta\eta$ is the difference between the hadronic final state and the scattered electron. 
\end{compactitem}

After the kinematic phase space selection and the above event criteria are enforced, the analysed dataset consists of 189,106 events.

\section{Simulations and theoretical calculations}
\label{sect:Theory}

Calculations based on the following Monte Carlo event generators and QCD calculations are compared to the data:

\begin{itemize}
  \item Djangoh~1.4~\cite{Charchula:1994kf,Kwiatkowski:1990es,Schuler:1991yg} uses Born-level matrix elements for NC DIS and dijet production, combined with the color dipole model from Ariadne~\cite{Lonnblad:1992tz} for higher-order emissions. Djangoh includes an interface to HERACLES~\cite{Kwiatkowski:1990es} for higher-order QED effects at the lepton vertex. The proton parton distribution function (PDF) used by Djangoh is CTEQ6L~\cite{Pumplin:2002vw}. Hadronization is simulated with the Lund hadronization
model~\cite{Andersson:1983ia,Sjostrand:1994kzr}, using parameters tuned to data by the ALEPH Collaboration~\cite{ALEPH:1996oqp}. 
 \item Rapgap~3.1~\cite{Jung:1993gf} implements Born-level matrix elements for NC DIS and dijet production
and uses the leading logarithmic approximation for parton showering. It is using the CTEQ6L PDF set and Lund hadronization implemented in Pythia \cite{Sjostrand:2001yu}.
  
\item The MC event generator Pythia~8.307~\cite{Sjostrand:2014zea,Pythia83} is used with two different parton-shower models: the default dipole-like $p_\perp$-ordered shower and the Dire~\cite{Hoche:2015sya,Hoche:2017iem,Hoche:2017hno} parton shower, which is an improved dipole shower with additional handling of collinear enhancements. Both implementations use the Pythia~8.3 default for hadronisation~\cite{Pythia83}, which is based on the Lund string model. The parton showers both use 0.118 for value of the strong coupling constant at the mass of the $Z$ boson, and both variations considered here use MMHT2014 as the hard PDF~\cite{Harland-Lang:2014zoa}.

\item A recent prediction from Ref.~\cite{Banfi:2023mhz} generalizes the POWHEG method to DIS, including handling of initial-state radiation off the lepton beam. This prediction, denoted Pythia+POWHEG, includes predictions in NLO QCD matched to parton showers using the POWHEG
method~\cite{Nason:2004rx,Frixione:2007vw}, which are then interfaced to Pythia~8.303. The default Pythia shower and hadronization schemes are applied. The Frixione-Kunszt-Signer (FKS) subtraction technique~\cite{Frixione:1995ms} is used to parameterize the radiation phase space.
  \item The MC event generator Herwig~7.2~\cite{Bellm:2015jjp} is
    also studied in three variants.
    For the default prediction, Herwig~7.2 implements leading-order
    matrix elements that are supplemented with an angular-ordered 
    parton shower~\cite{Gieseke:2003rz} and the cluster hadronization
    model~\cite{Webber:1983if,Marchesini:1991ch}.
    The second variant makes use of the MC@NLO method that implements NLO
    matrix element corrections. In addition, a matching with the default
    angular-ordered parton shower is performed~\cite{Platzer:2011bc}.
    The third variant also makes use of NLO matrix elements but applies the
    dipole merging technique and a dipole parton shower~\cite{Platzer:2011bc}. The PDF used for all three of the Herwig variations is MMHT2014~\cite{Harland-Lang:2014zoa}.
    The events generated with Herwig are further processed with Rivet~\cite{Bierlich:2019rhm}.
  \item Predictions are obtained with
    Sherpa~2~\cite{Sherpa:2019gpd,Gleisberg:2008ta}, where Comix~\cite{Duhr:2006iq} generates matrix elements for up to three final-state jets. The CKKW merging formalism~\cite{Catani:2001cc} is used to augment these jets with
    dipole showers~\cite{Catani:1996vz,Schumann:2007mg}, and
    the final-state partons are hadronized with cluster hadronization as implemented in
    AHADIC++~\cite{Winter:2003tt}. 
    As an alternative prediction, the hadronization step is performed
    with the Lund string fragmentation model~\cite{Sjostrand:2006za}. Both predictions use the Sherpa 2 default PDF, which is CT10~\cite{Lai:2010vv}.
    \item A set of predictions for the groomed 1-jettiness is provided by the Sherpa authors using a pre-release version of Sherpa 3. Sherpa 3 features a new cluster hadronization model, described in Ref.~\cite{Chahal:2022rid}.
     Matrix elements at NLO are 
     obtained from OpenLoops~\cite{Buccioni:2019sur}, and the resulting partons are showered via the Sherpa dipole
     shower~\cite{Schumann:2007mg} based on the truncated shower
     method described in Refs.~\cite{Hoeche:2009rj,Hoeche:2012yf}. Sherpa 3 additionally features intrinsic $k_{\mathrm{T}}$ of the partons within the proton, while the predictions from Sherpa 2 have no intrinsic $k_{\mathrm{T}}$ included. The model for partonic intrinsic $k_{\mathrm{T}}$ used in the Sherpa 3 prediction is a Gaussian form with mean of 0 GeV and  $\sigma=0.75$ GeV. The prediction has an associated uncertainty, defined as the extrema of a 7-point scale variation. 

\item  Another prediction at NNLO+NLL$^{\prime}$~\cite{Knobbe:2023ehi} is provided for the groomed 1-jettiness. The predictions are computed using the CAESAR formalism \cite{Banfi:2004yd} for NLL resummation as implemented in the Sherpa \cite{Sherpa:2019gpd} framework~\cite{Gerwick:2014gya, Baberuxki:2019ifp} and extended to cover the case of soft-drop groomed event shapes in Refs.~\cite{Marzani:2019evv, Baron:2020xoi}. The resummed results are matched to fixed order predictions at NNLO accuracy, derived with the techniques implemented in Sherpa in Ref. \cite{Hoche:2018gti}. Via a multiplicative matching, the calculation achieves NNLO+NLL$^\prime$ accuracy. A description of the full setup in the DIS case can be found in Ref.~\cite{Knobbe:2023ehi}. Non-perturbative corrections are applied via a transfer matrix approach \cite{Reichelt:2021svh} from Sherpa tuned to data from the LEP and HERA experiments~\cite{Chahal:2022rid,Knobbe:2023ehi}. The same techniques for calculating the groomed invariant mass are used here for the first time. Care is taken to evaluate logarithms of the form $\ln(M^2/Q^2)$ to avoid ambiguities arising from the normalisation of the observable. The calculation ultimately achieves the same NNLO+NLL$^\prime$ accuracy as before.

\item Predictions for the groomed invariant mass are provided using the formalism of SCET~\cite{Makris:2021drz}. Predictions are constructed at NNLL for the shape of the groomed invariant mass spectrum in the single-jet limit, corresponding to small values of the groomed invariant mass \Mgrsq $\ll$ \Qsq. The perturbative predictions are convoluted with a shape function to account for non-perturbative effects. The predictions are provided at two values of the shape function mean parameter, $\Omega_{\mathrm{NP}}$. No attempt is made to match the NNLL calculation to a fixed-order prediction. 

\end{itemize}

\section{Observables}
\label{sect:Analysis}

\subsection{Breit frame}
\label{sect:BreitFrame}

The reported measurements are carried out in the Breit frame of reference~\cite{Streng:1979pv}. The Breit frame is the reference frame in which 

\begin{equation}
2\xB\Vec{P}+\Vec{q} = 0,
\label{eq:BreitFrame}
\end{equation}

\noindent
where $\Vec{P}$ is the three-momentum of the incoming proton beam and $\Vec{q}$ is the three-momentum of the exchanged boson. As for the laboratory frame of reference, we choose the positive $z$ axis of the Breit frame to be the direction of the incoming proton. In the quark-parton model, the Breit frame is the frame of reference in which the struck parton is initially aligned with the $z$ axis with $p_z = Q/2$ and leaves along the same axis with $p_z = -Q/2$ after the momentum transfer from the space-like virtual boson. The maximum available longitudinal momentum in the Breit frame is therefore $Q/2$, and at Born level the scattered parton has transverse momentum $\pT=0$. 

The direction of the proton remnant is defined as $\etaB = +\infty$, while the direction of the struck parton is defined as $\etaB = -\infty$. The region $\etaB > 0$ is denoted the remnant hemisphere (RH), and the $\etaB < 0$ region is denoted the current hemisphere (CH). Events with large \pT\ in the Breit frame therefore correspond to multi-jet topologies, e.g. QCD Compton scattering in which the quark recoils with significant \pT from a hard gluon emission. For the kinematic selection used in this analysis, namely $Q^2 > 150$ GeV$^2$ and inelasticity $ 0.2 < y < 0.7$, the scattered electron is typically at mid-rapidity ($\etaB\sim0$) in the Breit frame. This phase space also corresponds to the region in which the magnitude of the Lorentz boost from the lab frame to the Breit frame is small, thus minimizing the event-to-event change in the detector acceptance in the Breit frame~\cite{Thompson:1993cg}.

\subsection{Jet and event clustering in the Breit frame: Centauro algorithm}
\label{sect:Centauro}

In the HERA convention, the leading-order quark-parton model process $eq\rightarrow eq$ produces a jet at $\etaB=-\infty$ in the Breit frame. Such jets will not be captured by longitudinally-invariant \kT-type sequential recombination clustering algorithms applied in that frame~\cite{Thompson:1993cg,Arratia:2020ssx}, since those algorithms cluster based on the \emph{transverse} component of particle momenta that is largely boosted away by the transformation to the Breit frame. Additionally, the distance $d_{ij}$ between particles becomes large in the direction of the struck parton due to the factor $R^2=\sqrt{\Delta \eta_{\mathrm{Breit}}^2+\Delta\phi^2}$ in the denominator of the distance metric for those algorithms. 


Centauro~\cite{Arratia:2020ssx} is a sequential-recombination jet algorithm that overcomes these limitations by means of an asymmetric clustering measure, which preferentially clusters a jet from radiation in the current hemisphere of the Breit frame. Centauro thus clusters the Born-level configuration into a jet more readily than other laboratory and Breit frame algorithms. The Centauro distance measure is
\begin{equation}
d_{ij} = (\Delta\Bar{\eta}_{ij})^2+2\Bar{\eta}_i\Bar{\eta}_j(1-\cos(\Delta\phi_{ij})),
\label{eq:CentauroDistance}
\end{equation}
where $\Delta\phi_{ij}$ is the azimuthal angle difference measured in the Breit frame between pairs of objects that are candidates for clustering, and $\Delta\Bar{\eta}_{ij}=\Bar{\eta}_i-\Bar{\eta}_j$ is the difference in an angular variables measuring along the longitudinal direction of the Breit frame. The angular variable is defined as
\begin{equation}
\Bar{\eta}_i = -\frac{2p_{\mathrm{T},i}}{n\cdot p_{i}},
\label{eq:CentauroEtabar}
\end{equation}
\noindent
where $p_i$ is the four-momentum of particle $i$, $p_{\mathrm{T},i}$ is its transverse momentum in the Breit frame, and $n$ is the four-vector of the beam proton, normalized to unity energy.

The Centauro algorithm can be used as a traditional jet-finding algorithm to reconstruct jets, but it can also be applied to the entire DIS event (equivalent to setting the jet radius $R\rightarrow\infty$) to generate a clustering tree in which the last clustered radiation is farthest from the nominal struck parton direction in the Breit frame. As discussed in Ref.~\cite{Makris:2021drz}, the natural quantity for comparison of branches in this tree is the Lorentz-invariant momentum fraction \zi, \[z_i=\frac{P\cdot p_i}{P\cdot q},\] which in the Breit frame represents the fraction of the virtual boson momentum carried by the object $i$.  Branches of the tree with low \zi\ are either soft or at wide angles with respect to the virtual boson. Branches with high \zi\ are likely to be fragments of the struck parton. 

\subsection{Event grooming}
\label{sect:Grooming}

Event grooming follows the procedure described in Ref.~\cite{Makris:2021drz}, as follows. All four-vectors in the event are clustered into a tree by the Centauro algorithm; the tree is iteratively declustered in order reverse to the initial clustering; and at each declustering step the values of \zi\ of the branches are compared to the grooming condition,

\begin{equation}
\frac{\text{min(}\zi,\zj)}{\zi+\zj} > \zcut.
\label{eq:GroomingCut}
\end{equation}

\noindent
If the grooming condition is not met, the branch with smaller \zi\ is removed and the remaining branch is again subdivided and compared to the grooming condition. The procedure continues until the grooming condition is met. Events in which the algorithm queries the full clustering tree without the grooming condition being met are removed from the analysis and do not contribute to the measured event shape cross section. The fraction of events which do not pass the grooming condition naturally increases with $z_{\mathrm{cut}}$. This approach is a version of the modified MassDrop Tagger (mMDT) grooming algorithm with $\beta=0$~\cite{Marzani:2017mva}, adapted for DIS with \zi\ playing the role of $p_{\mathrm{T}_i}$ in standard mMDT. 

\begin{figure*}[t]
\centering
\includegraphics[scale=0.35]{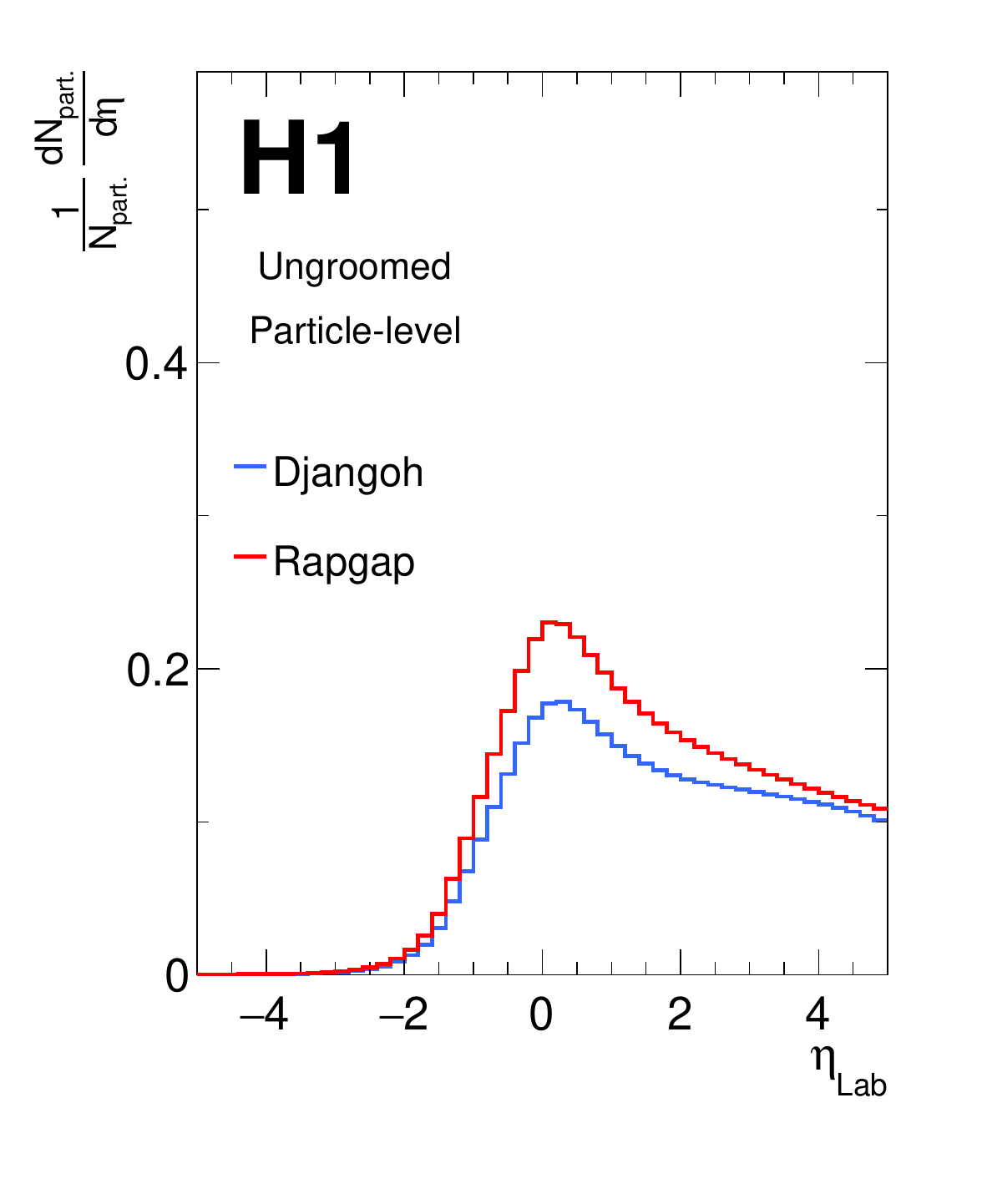}
\includegraphics[scale=0.35]{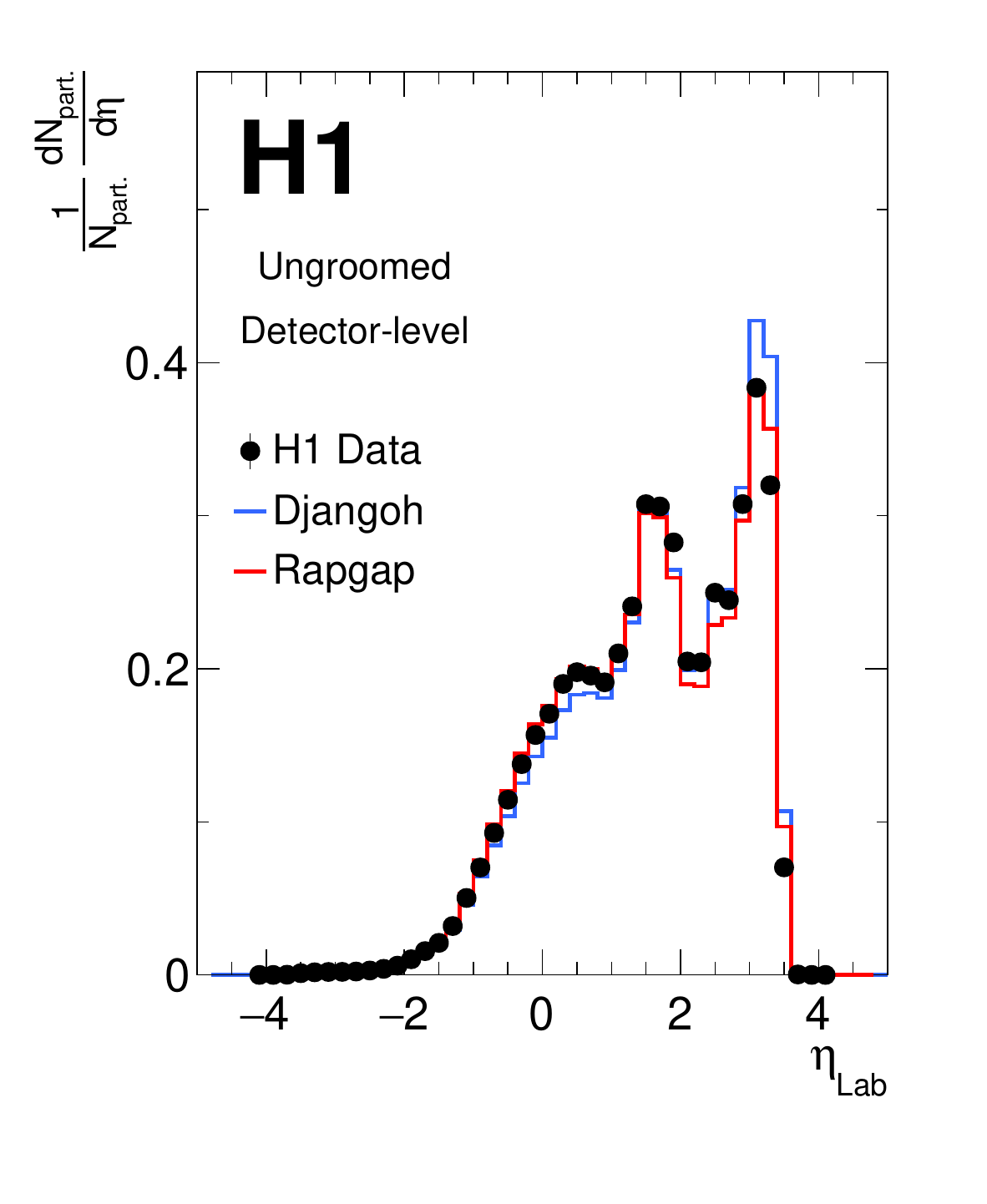}
\includegraphics[scale=0.35]{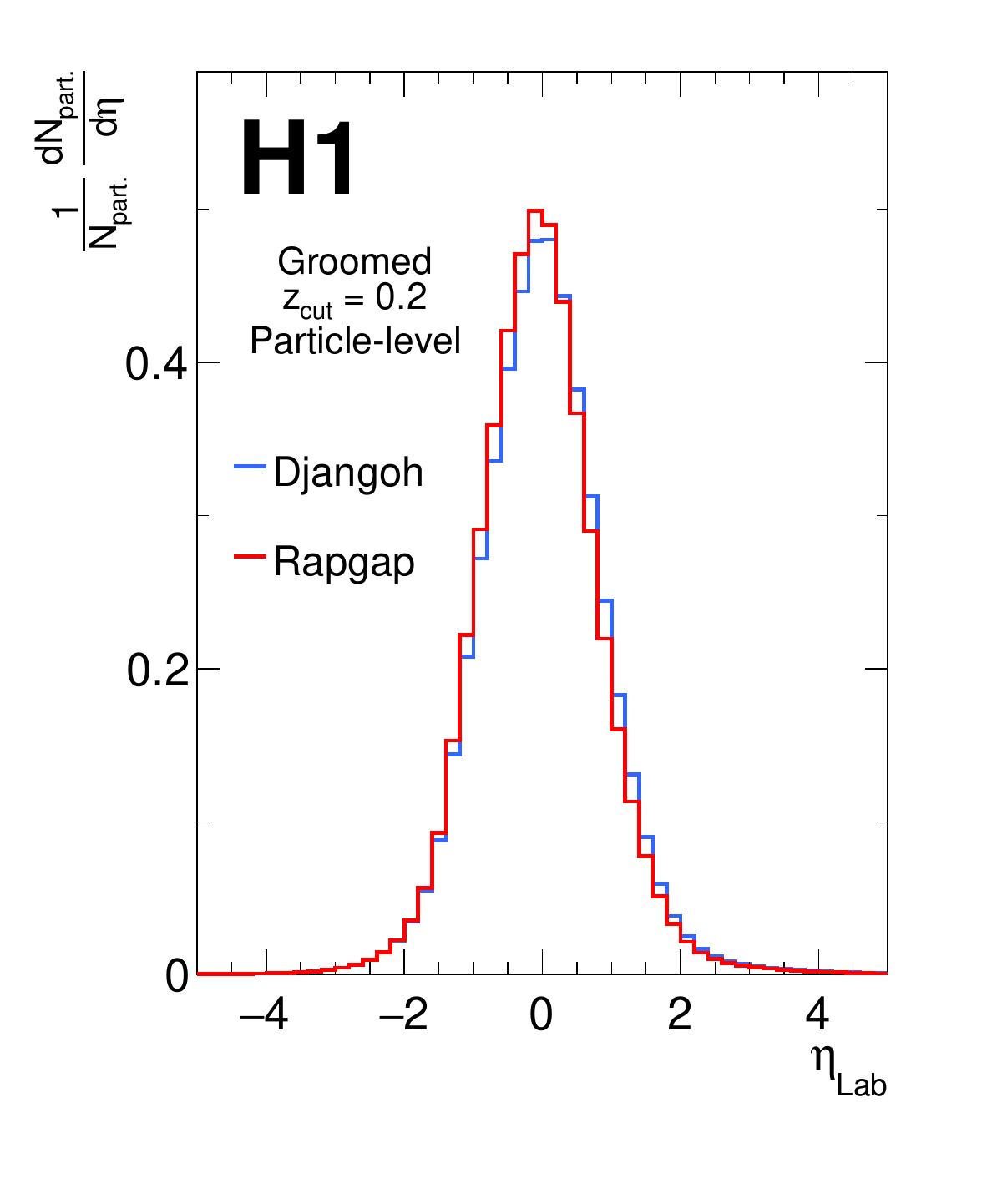}
\includegraphics[scale=0.35]{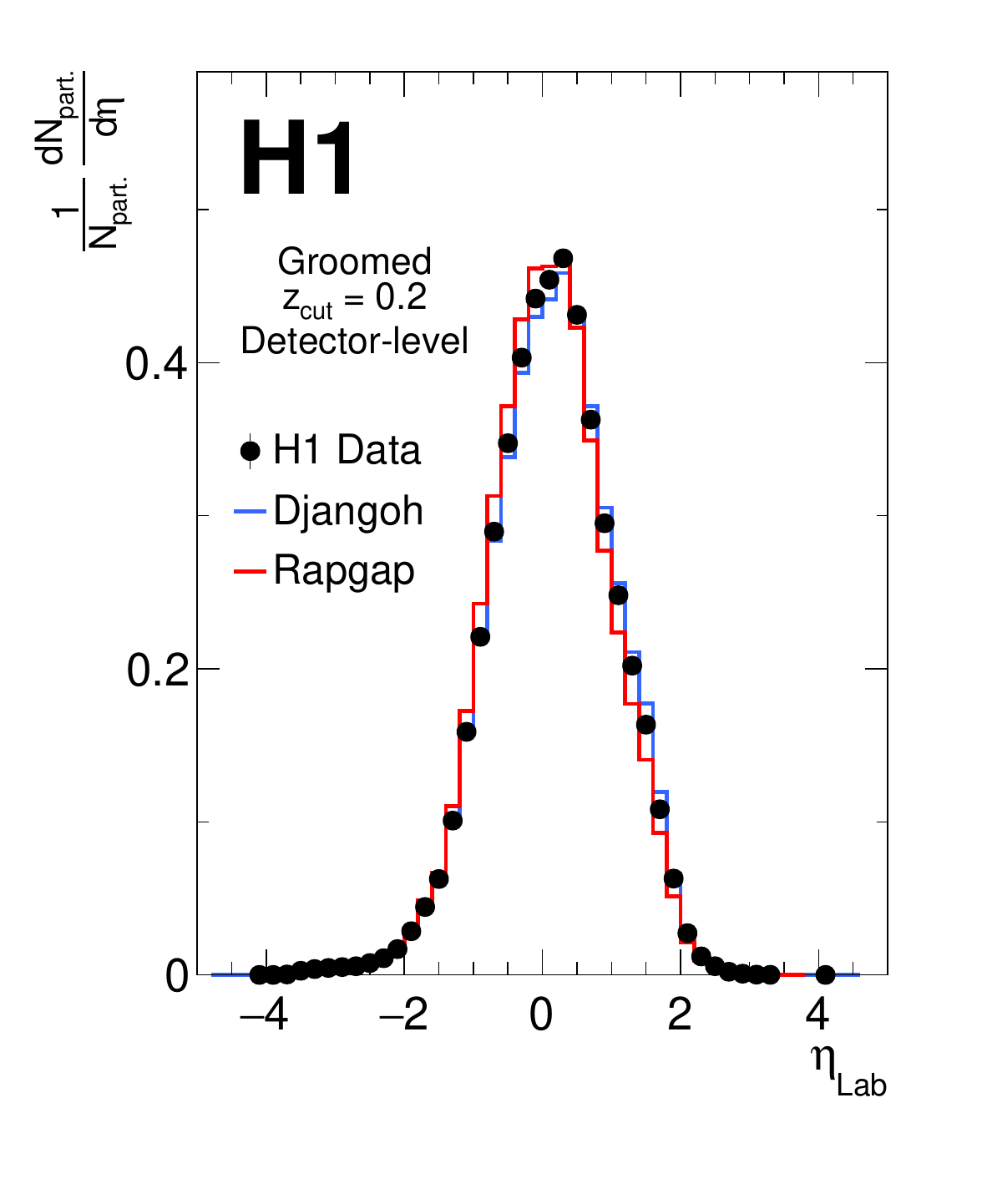}
\caption{Normalized distribution of particles measured in data and simulated by Djangoh and Rapgap, as a function of pseudorapidity in the laboratory frame. Left: particle-level; right: detector-level. Upper: ungroomed (\zcut=0); lower: groomed with \zcut=0.2.}
\label{fig:LabEtasA}
\end{figure*}

Figure~\ref{fig:LabEtasA} shows single-particle pseudorapidity distributions for groomed and ungroomed events at both particle- and detector-level. ``Particle-level" here refers to the quantities produced by the event generator, where the final state comprises particles with proper lifetime greater than 8 ns. ``Detector-level" refers to the four-vectors reconstructed after the particles from the generator are passed through a GEANT~\cite{Brun:1082634} detector simulation program, as well as several reconstruction algorithms. The complex shape of the ungroomed detector-level pseudorapidity distribution in the top right panel of Fig.~\ref{fig:LabEtasA} can be attributed to the transition between the barrel and forward sections of the LAr, as well as the contribution of secondary interactions in the detector and beamline. 

Figure~\ref{fig:LabEtasA} also compares the detector-level simulated distributions to raw data, with good agreement found. For the data, the fraction of accepted events that pass the grooming cut is 99.3\% for \zcut = 0.05, 98.4\% for \zcut = 0.1, and 92.7\% for \zcut = 0.2.

\subsection{Groomed event shapes}
\label{Sec:ESDefs}

Classical global event shape observables incorporate a summation over all particles in an event, including those which are produced at small angles with respect to the beam. Monte Carlo simulations show that, for high-\Qsq\ \ep\ DIS events such as those in this analysis, 30--40\% of generated particles fall in the forward region $\etaL>3.5$, beyond the detector acceptance. In previous event shape analyses at HERA~\cite{Adloff:1997gq,Adloff:1999gn,Aktas:2005tz,ZEUS:2002tyf,ZEUS:2006vwm}, the impact of the missing forward acceptance was reduced by limiting the measurement to particles in the current hemisphere, which are better contained by the detector. In that case, however, theoretical calculations of event shapes must also include radiation that is nominally emitted in the remnant hemisphere but enters the current hemisphere at higher order, generating non-global logarithms that can compromise theoretical precision~\cite{Kang:2013nha,Makris:2021drz}. These limitations motivate consideration of methods that can ameliorate both the experimental and theoretical challenges typically associated with event shape observables.

It can be seen from Fig.~\ref{fig:LabEtasA} that the application of grooming alleviates the need for a redefinition of event shape observables, since the grooming procedure grooms away the particles produced outside the detector acceptance. This follows from the fact that at high \Qsq, the exchanged virtual boson typically points toward the central region of the detector. Since the grooming tends to remove particles that are anti-collinear and at wide angles with respect to the exchanged boson, the particles that survive the grooming are generally well-contained within the central region of the detector. For the hardest grooming cut considered here, $\zcut\ = 0.2$, only 0.5\% of particles surviving the grooming at particle-level are beyond the forward acceptance of the detector. QCD initial-state radiation, beam remnants, and wide-angle soft radiation are largely groomed away. The remaining particles therefore consist predominantly of fragments of the struck parton, which are collimated in the virtual boson direction. Groomed event shape observables are calculated from these surviving particles. 

This paper reports two groomed event shape observables: \Mgrsq, the groomed invariant mass, and \tauonebgr, the groomed 1-jettiness. The observable \Mgrsq\ is defined as
\begin{equation}
\Mgrsq = \left(\sum_{i\in \mathrm{groomed}(\zcut)}p_i\right)^{2},
\label{eq:Minvgr}
\end{equation}
\noindent
where the sum over $i$ runs over all particles in the hadronic final state that survive the grooming for the specified value of \zcut, and $p_i$ is the four-momentum of particle $i$. For comparison with the predictions in Ref.~\cite{Makris:2021drz}, the groomed invariant mass is expressed as the natural logarithm of \Mgrsq normalized to \Qminsq, the minimum value of \Qsq in the event population; i.e. 
\begin{equation}
\text{GIM}\equiv\GIMln.
\label{eq:GIMnormalized}
\end{equation}
Measurements of the single- and double-differential GIM cross sections are reported in the present paper. The double-differential cross sections are presented as functions of \Qsq. For the single-differential measurement, as well as the normalized double-differential measurement presented in Fig.~\ref{fig:UniversalityGIM2}, \Qminsq\ is fixed at 150 \GeVsq. For the double-differential cross section measurement presented in Fig.~\ref{fig:2DGIM}, \Qminsq\ is set to the lower edge of each \Qsq bin.

The 1-jettiness event shape observable $\tau_1$ and its variants, $\tau^{a}_{1}$, $\tau^{b}_{1}$, and $\tau^{c}_{1}$, are defined in Ref.~\cite{Kang:2013nha}. The 1-jettiness variant \tauoneb\ is chosen for this analysis because it aligns best with the grooming procedure: both the Centauro clustering algorithm and the grooming procedure are defined in the Breit frame, which is also the natural frame for \tauoneb. In contrast, $\tau^{a}_{1}$ uses a jet found in the lab frame, and $\tau^{c}_{1}$ uses the center-of-mass frame. The groomed observable \tauonebgr\ is defined as
\begin{equation}
\tauonebgr=\frac{2}{Q^2} \sum_{i\in \mathrm{groomed}(\zcut)}\text{min(}q_B\cdot p_i,q_J\cdot p_i),
\label{eq:tauonebgr}
\end{equation}
\noindent
where $q_{B}=xP$ and $q_{J}=xP+q$, and the sum likewise runs over all the hadronic final state particles that survive the grooming procedure for the chosen value of \zcut.

Equation~\ref{eq:tauonebgr} projects each particle 4-vector onto both $q_J$, the virtual boson 4-vector, and $q_B$, the beam 4-vector, and selects the axis to which it is best aligned. Particles in the current hemisphere are better aligned with $q_J$, while particles in the remnant hemisphere are better aligned with $q_B$. The observable \tauonebgr\ takes values between 0 and 1, with values near 0 corresponding to collimated events resembling a single jet and values near 1 corresponding to multi-jet events. If momentum conservation in the Breit frame is assumed, $\tau_1^b$ is formally equivalent to the DIS thrust normalized by $Q$~\cite{Kang:2013nha}. 
The sum is normalized by the value of \Qsq measured in the corresponding event.

Since radiation in the proton-going direction is removed by grooming, groomed event shapes are more tightly correlated with the struck parton direction than standard DIS event shapes~\cite{Dasgupta:1998xt,Dasgupta:1997ex,Dasgupta:2001sh,Dokshitzer:1998pt}. At leading order, groomed events can therefore be considered as jets; grooming effectively defines a jet without imposing a jet radius cutoff for the clustering. 
\subsection{\boldmath Choice of \zcut~value\unboldmath}
\label{Sec:zcut_choice}
In the analytic SCET calculations, the value of \zcut\ should be chosen to respect the factorization of the calculation~\cite{Makris:2021drz}. Two regions are defined in the calculation: $ 1 \gg \zcut \gg \GIM$, and $1 \gg \zcut \simeq \GIM$. The reach in \Qsq is fixed primarily by the integrated luminosity and center-of-mass energy, and the reach to low masses is limited by the detector resolution for small angles in the Breit frame. Values of \zcut ~greater than 0.3 not only begin to violate the condition $\zcut \ll 1$, they also result in a large fraction of events with small groomed invariant mass that are challenging to reconstruct precisely. Therefore, in this analysis we report event shape distributions for \zcut\ values of 0.05, 0.1, and 0.2.


\section{Corrections}
\label{sect:Corrections}

The corrected differential cross section in a bin of event shape observable $e$ is defined as 

\begin{equation}
  \frac{d\sigma_i}{de} = \frac{\sum_j A^+_{ij}(n_\text{data,j} - n_\text{Bkg,j})} {\Lint \cdot
    \Delta}\cdot c_{\text{QED,i}}\,,
\label{eq:XsectionCorrection}
\end{equation}

\noindent
where the indices $i$ and $j$ represent particle-level and detector-level quantities, respectively; $\Delta$ is the bin width; $A^+_{ij}$ is the regularized inverse of the detector response matrix; \Lint\ is the analysed integrated luminosity; $n_\text{data,j}$ is the number of events measured in bin $j$; $n_\text{Bkg,j}$ is the number of estimated background events in bin $j$; and $c_{\mathrm{QED}}$ is the QED correction factor. 

 The data are corrected to the non-radiative particle-level in the phase space of $\Qsq> 150~\GeVsq$~and $0.2<y<0.7$. Detector effects are corrected by regularized unfolding using the TUnfold package~\cite{Schmitt:2012kp}, and QED effects are corrected bin-by-bin. No hadronization corrections are applied to the data.
 
\subsection{Regularized unfolding}
\label{sect:Unfolding}

TUnfold utilizes a least-squares fit technique with Tikhonov regularization. The so-called ``curvature'' mode of TUnfold is utilized, which regularizes the second derivative of the output distribution. The regularization is performed at values of the regularization parameter $\tau$ that minimize the influence of the unfolding on the final result while maintaining good closure.

The detector response matrix for unfolding is calculated as the average of the respective matrices built using events generated by Rapgap~\cite{Jung:1993gf} and Djangoh~\cite{Charchula:1994kf,Kwiatkowski:1990es,Schuler:1991yg}. The simulated datasets correspond to an integrated luminosity of $\Lint= 40$ fb$^{-1}$ for both Djangoh and Rapgap. The generated events are then passed through the H1 detector simulation implemented in GEANT3\cite{Brun:1082634} and augmented with a fast calorimeter simulation~\cite{Kuhlen:1992ey,Glazov:2010zza}. The same reconstruction algorithms that are used for data are applied to the output of the simulation. The response matrix has three bins in the reconstructed observable for each bin of measured data. 

The following models are used to evaluate the number of background events measured in each bin, $n_\text{Bkg,j}$:
\begin{compactitem}
\item NC DIS is simulated by Djangoh\cite{Charchula:1994kf} and Rapgap\cite{Jung:1993gf}. For $60 < Q^2 < 150~\mathrm{GeV}^2$, Djangoh and Rapgap are used, while for $4 < Q^2 < 60~\mathrm{GeV}^2$, only Djangoh is used. The contribution of these backgrounds to the measured event sample is around 5\% and is dominated by migration to higher \Qsq of events near to the kinematic boundary, i.e.\ $Q^2 < 150$~\GeVsq.
\item Events with $Q^2 < 4$ GeV$^2$, including photoproduction, are simulated by Pythia~6.2~\cite{Sjostrand:1993yb,Sjostrand:2001yu}.
\item QED Compton scattering is simulated by COMPTON~\cite{Courau:1992ht}.
\item Di-lepton production is simulated by GRAPE\cite{Abe:2000cv}.
\item Deeply virtual Compton scattering is simulated by MILOU\cite{Perez:2004ig}.
\item Charged-current DIS is simulated by Djangoh.
\end{compactitem}
The contributions of all sources of background other than NC DIS are negligible.

Closure of the unfolding procedure is tested by unfolding the detector-level distribution as determined via Rapgap with the response matrix generated by Djangoh, and vice versa. The output distribution of the unfolding procedure is compared to the corresponding particle-level distribution to determine whether the procedure is returning results close to the truth. Typical values of $\tau$ are $1\cdot10^{-5}$ and $4\cdot10^{-4}$ for the single- and double-differential distributions, respectively. 

\subsection{QED radiation and corrections}

The radiation of photons off the electron affects the cross section in several ways. Initial-state emission of a real photon off the electron distorts the measurement of $Q^2$, $y$, and $x$, occasionally producing an energetic cluster in the SpaCal. Final-state radiation is typically emitted collinear to the scattered electron and thus produces one energetic cluster in the calorimeter, but occasionally the photon will be produced at a larger angle with respect to the electron and will be resolved. In both cases, these photons must be removed from the hadronic final state since they tend to lie at mid-rapidity in the Breit frame and therefore can significantly disturb the grooming procedure. Additionally, virtual corrections to the NC DIS process can change the overall normalization and shape of the inclusive cross section. 
 
QED effects are included in Djangoh and Rapgap via an interface to the HERACLES program~\cite{Kwiatkowski:1990es}. HERACLES simulates the first-order electroweak corrections to both $e^+p$ and $e^-p$ DIS, including virtual corrections and real photon emission from the lepton. The data are corrected for these effects by applying a bin-by-bin factor, $c_{\mathrm{QED}}$, which is defined as the ratio between the non-radiative and radiative particle-level distributions. The data, which are a mixture of $e^+p$ and $e^-p$ collisions, are corrected to the $e^-p$ cross section. This effect is also encapsulated in $c_{\mathrm{QED}}$. The value of $c_{\mathrm{QED}}$ is similar for both observables and has a uniform value of about 1.15. The magnitude of the QED correction is a result of the cut on $E-p_z$ being applied on the radiative particle-level. A non-radiative particle-level event always has $E-p_z = 2E_e$, whereas the radiative particle-level event has been defined with the ISR photon excluded, such that $E-p_z = 2(E_e-E_{\gamma})$, where $E_{\gamma}$ is the energy of the photon radiated by the electron in the initial state. The result is that in the radiative particle-level, the cross section is decreased by the likelihood that $2(E_e-E_{\gamma}) < 50~\GeV$, which is around 15\% in the kinematics of this measurement. The values of $c_{\mathrm{QED}}$ are presented in the data tables in the appendix. In the highest $Q^2$ bin of the double-differential measurement, $c_{\mathrm{QED}}$ has values around 20\%, due to the difference between the $e^-p$ and $e^+p$ cross sections at high $Q^2$.

%
%

\section{Uncertainties}
\label{sect:Uncertainties}

The following components of the analysis contribute to the systematic uncertainty of the reported cross sections. All sources of uncertainty are evaluated with both Rapgap and Djangoh. The average of the uncertainties as determined using the two models is used as the uncertainty on the data. The total systematic uncertainty is defined as the sum in quadrature of the individual systematic uncertainties arising from the sources described below. 

\subsection{Alignment}

The polar angle alignment of the tracking detectors with the liquid argon calorimeter has a precision of 1 mrad~\cite{H1:2012qti}. This precision results in an uncertainty in the measured position for all HFS objects and for the scattered electron. The HFS and electron polar angle uncertainties are considered separately, and each is passed through the unfolding procedure. The resulting uncertainties in the final distributions are typically $\sim 1\%$. The values reported in the data tables in Sec.~\ref{sect:Tables} are signed quantities corresponding to the difference between the nominal angles and the angles after the systematic shift of the polar angle of all simulated objects upwards by 1 mrad.

\subsection{Energy scales}

The measured energy of the scattered electron has a precision of 0.5\,\% in the backward and central regions of the detector and a precision of $1\,\%$ in the forward region~\cite{H1:2012qti}. The uncertainty in final distributions due to this precision is determined by varying the scattered electron energy and passing the modified events through the unfolding procedure. The resulting uncertainty has a value of less than $\sim3\%$.

Independent cluster energy calibrations are used to describe clusters inside and outside of high-\pT\ jets~\cite{Kogler:2011zz,H1:2014cbm}. The uncertainty in energy scale of particles inside of high-\pT\ jets is denoted as the jet energy scale uncertainty (JES), and the uncertainty in energy scale of particles outside of jets is denoted as the residual cluster energy scale uncertainty (RCES). These energy scales are independently varied by a factor of $1\%$, and the resulting distributions are passed through the unfolding procedure. The difference in unfolded distributions between the variations and nominal scale values provides the corresponding uncertainty, which is typically 1--2\%. 

The value of these uncertainties as reported in the data tables is the signed difference between the unfolded result with the nominal energy scale and the energy scale shifted upwards, i.e. a negative value corresponds to the case where the value in a given bin was increased after the systematic shift.

\subsection{Integrated luminosity and normalization effects}

The uncertainty in integrated luminosity is 2.7\%~\cite{H1:2012wor}, which is applied to the final cross sections. This uncertainty additionally accounts for several other small normalization uncertainties, including trigger efficiency, the QED correction, the calorimeter noise suppression algorithm, and electron identification.

\subsection{Unfolding}

Three sources of uncertainty from the unfolding procedure are studied. In almost all bins of the measurement, unfolding-related uncertainties dominate over detector-related uncertainties.

\paragraph{Model Dependence:} The uncertainty from the model dependence of the unfolding is estimated as half the difference between the spectra unfolded using the migration matrices from Rapgap and Djangoh, respectively. This is the dominant systematic uncertainty in many bins of the measurement, with typical values of 5--10\% for the single-differential.

\paragraph{Regularization:} The uncertainty associated with regularization is determined by varying the regularization parameter by a factor of two larger and smaller than its nominal value. For the single-differential cross sections, the regularization uncertainty is similar in size to the model dependence uncertainty.
 
\paragraph{Statistics:} The statistical uncertainty is determined by a resampling procedure, in which the input data to the unfolding are varied according to the statistical precision associated with the number of events in each bin and then unfolded. For each observable, this procedure was repeated one thousand times. In each bin of the measurement, the standard deviation of the one thousand replicas is taken as an uncertainty on the value of the bin. This source of uncertainty is typically sub-leading, excepting a few bins with limited statistics in the double-differential distributions. 



\section{Results}
\label{sect:Results}

In this section we present cross sections of the normalized groomed invariant mass GIM Eq.~\eqref{eq:GIMnormalized} and groomed 1-jettiness \tb Eq.~\eqref{eq:tauonebgr}, fully corrected for detector and QED effects as described in Section~\ref{sect:Corrections}. The analysis phase space is defined by $0.2 < y < 0.7$ and $Q^2 > 150~$\GeVsq. Section~\ref{sect:Theory} describes the MC models and analytic pQCD calculations that are compared to the data. The data tables are provided in Section~\ref{sect:Tables}.


\subsection{Single-differential cross sections}
\label{sect:SingleMC}


\begin{figure*}[t]
 \begin{subfigure}[b]{\textwidth}
    \centering
    \includegraphics[width=0.88\textwidth]{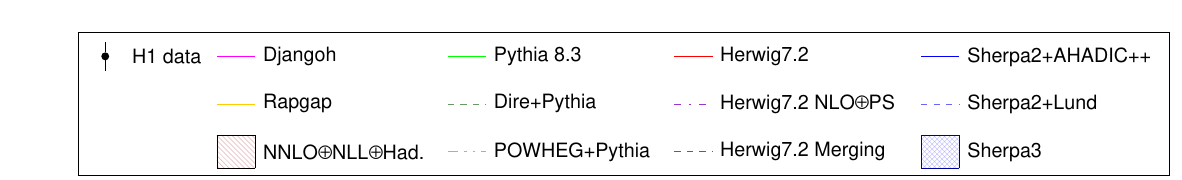}
\end{subfigure}
  \centering
   \begin{overpic}[trim={0.9cm 0 1cm 0},clip,scale=0.4]{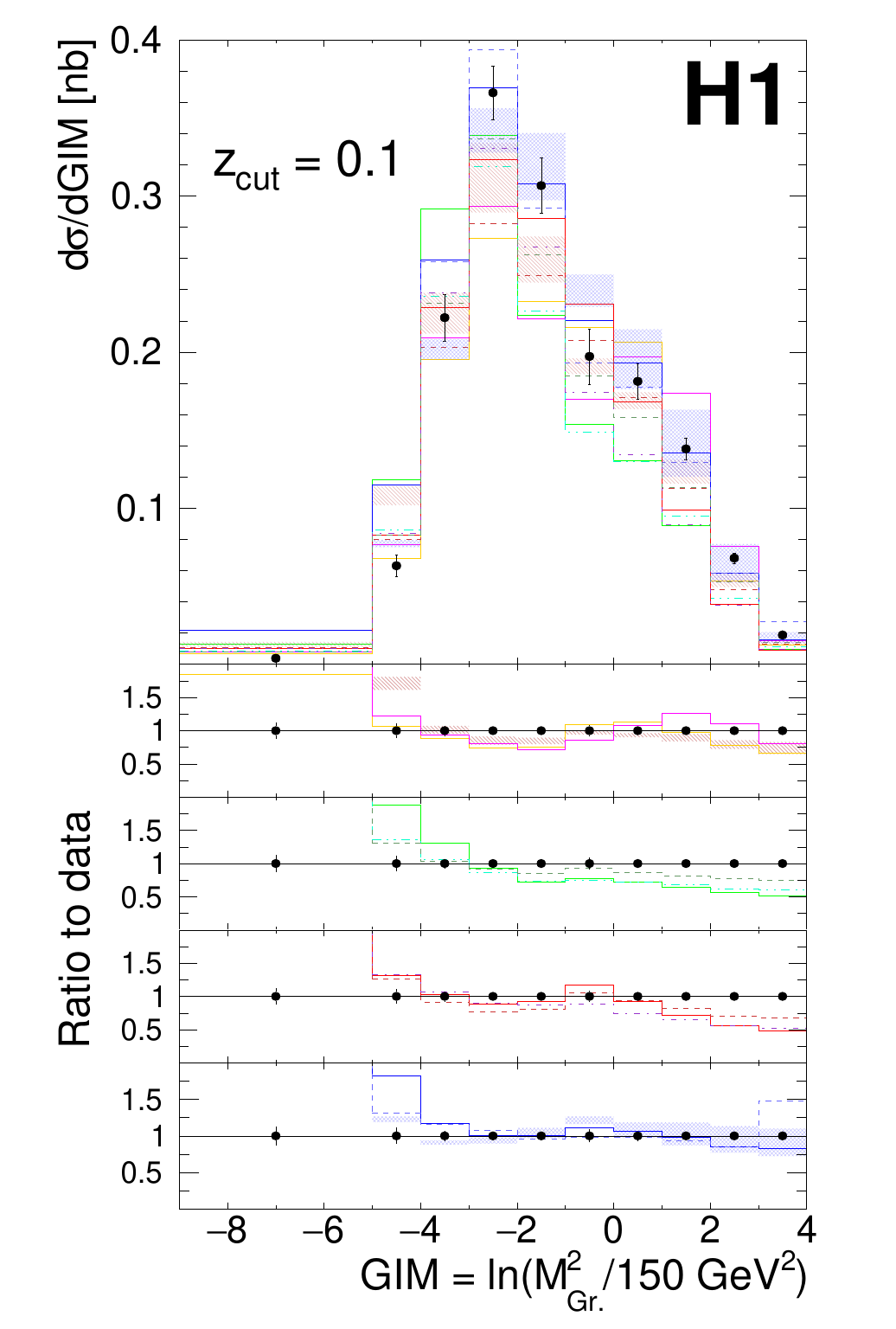}
     \put(60,0){\includegraphics[trim={3.5cm 0 1cm 0},clip,scale=0.4]{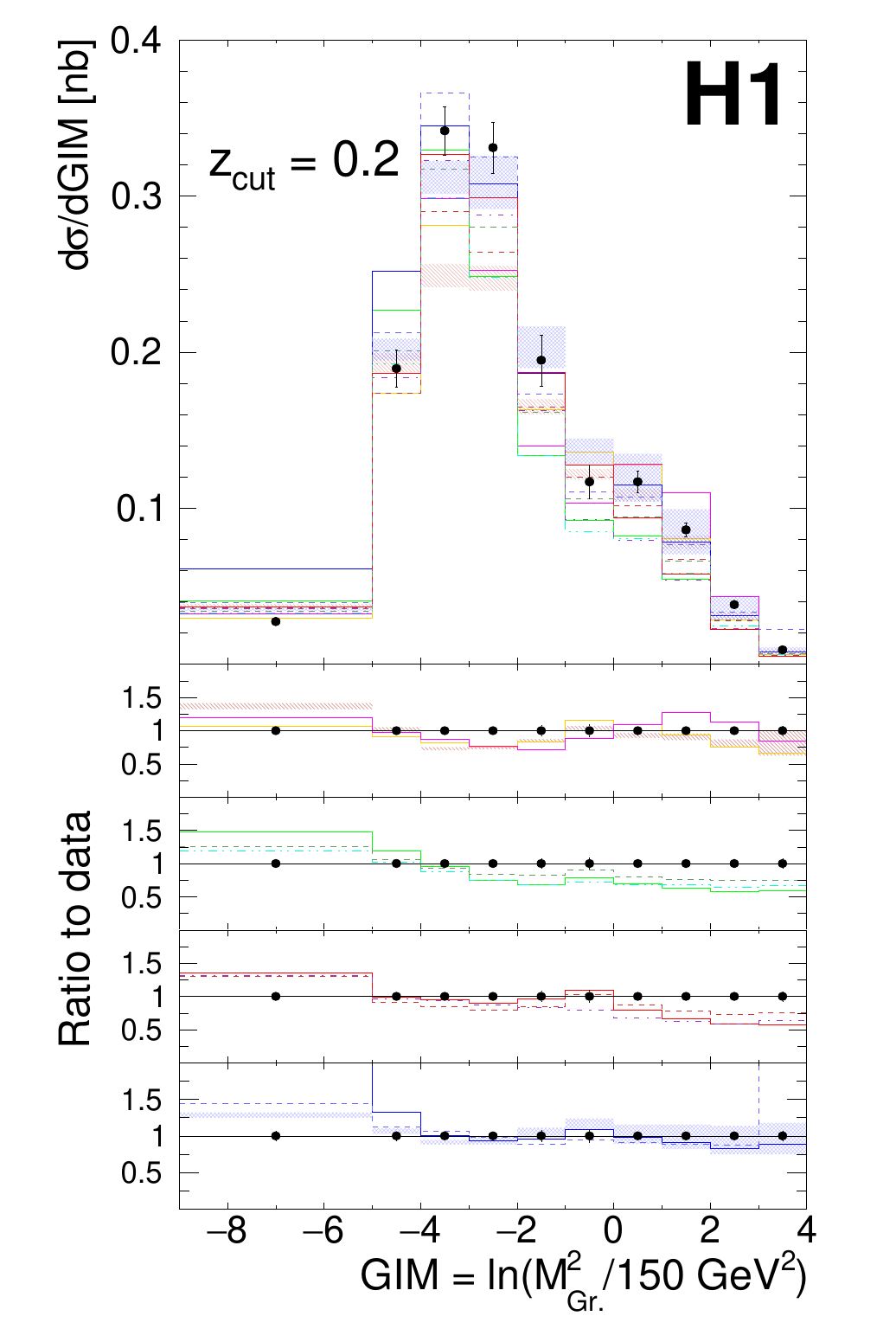}}
     \put(-50,0){\includegraphics[trim={1cm 0 1cm 0},clip,scale=0.4]{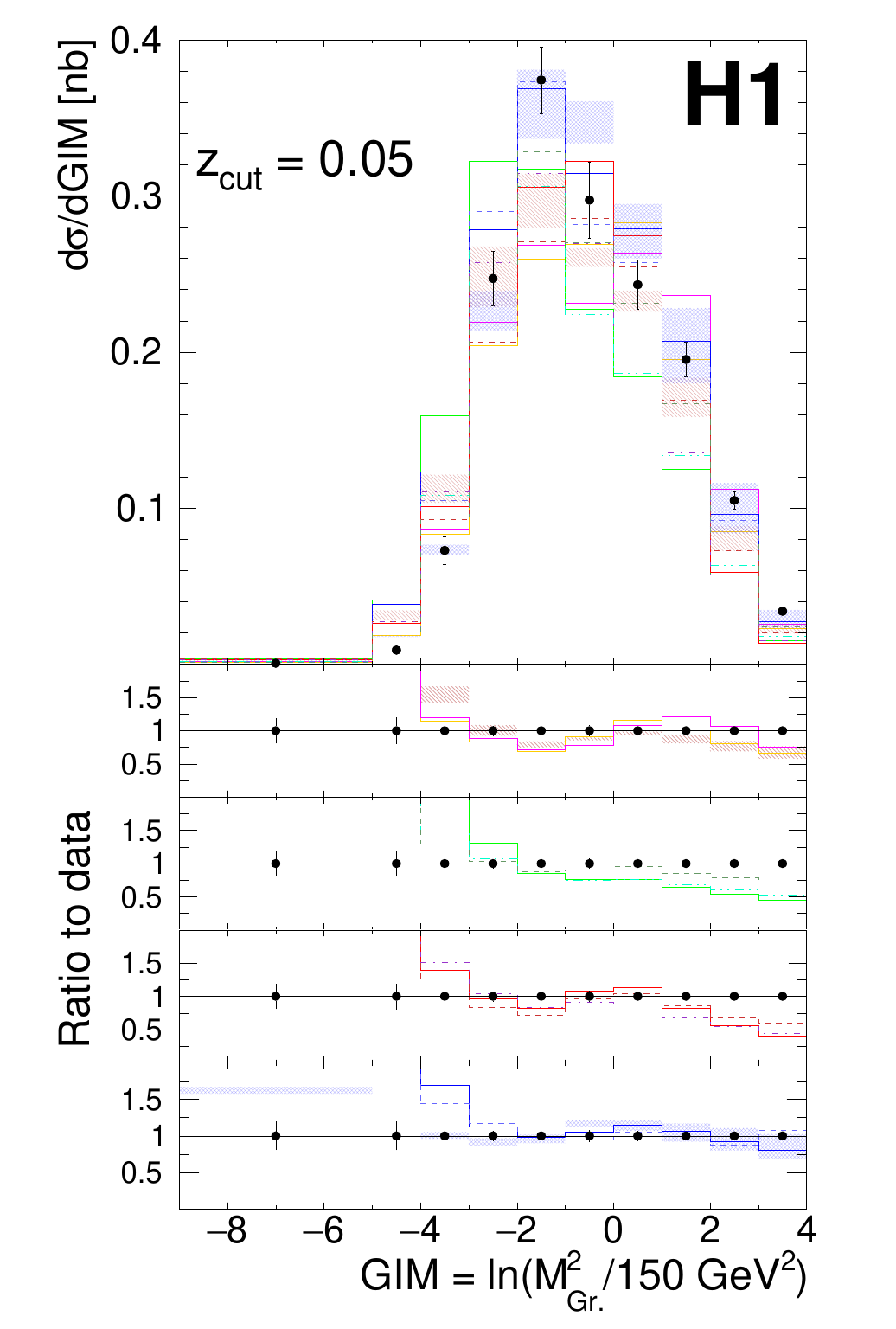}}
  \end{overpic}
\caption{Differential cross section of groomed invariant mass GIM~$=\GIMlnone$ in \ep\ DIS at $\sqrts=319~ \mathrm{GeV}$, for $\zcut=0.05$, 0.1 and, 0.2. The value of \Qminsq is set to 150 \GeVsq. The phase space is restricted to $\Qsq>150$~\GeVsq~and $0.2 < y < 0.7$. Uncertainty bars on the data show the quadrature sum of the statistical error and systematic uncertainty. The Monte Carlo event generators and pQCD calculations that are compared to the data are described in Section~\ref{sect:Theory}.}
\label{fig:MC1DGIMs}
\end{figure*}

\begin{figure*}[t]
\begin{subfigure}[b]{\textwidth}
    \centering
    \includegraphics[width=0.88\textwidth]{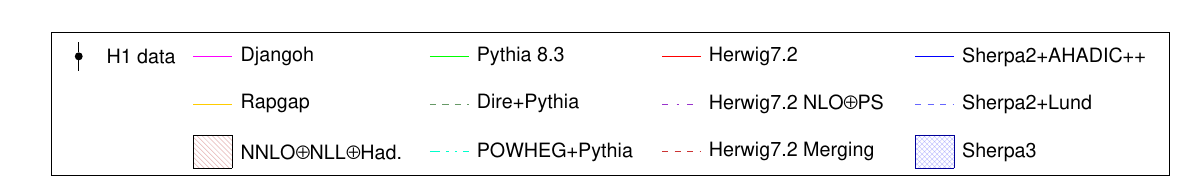}
\end{subfigure}
  \centering
   \begin{overpic}[trim={0.9cm 0 0 0},clip,scale=0.4]{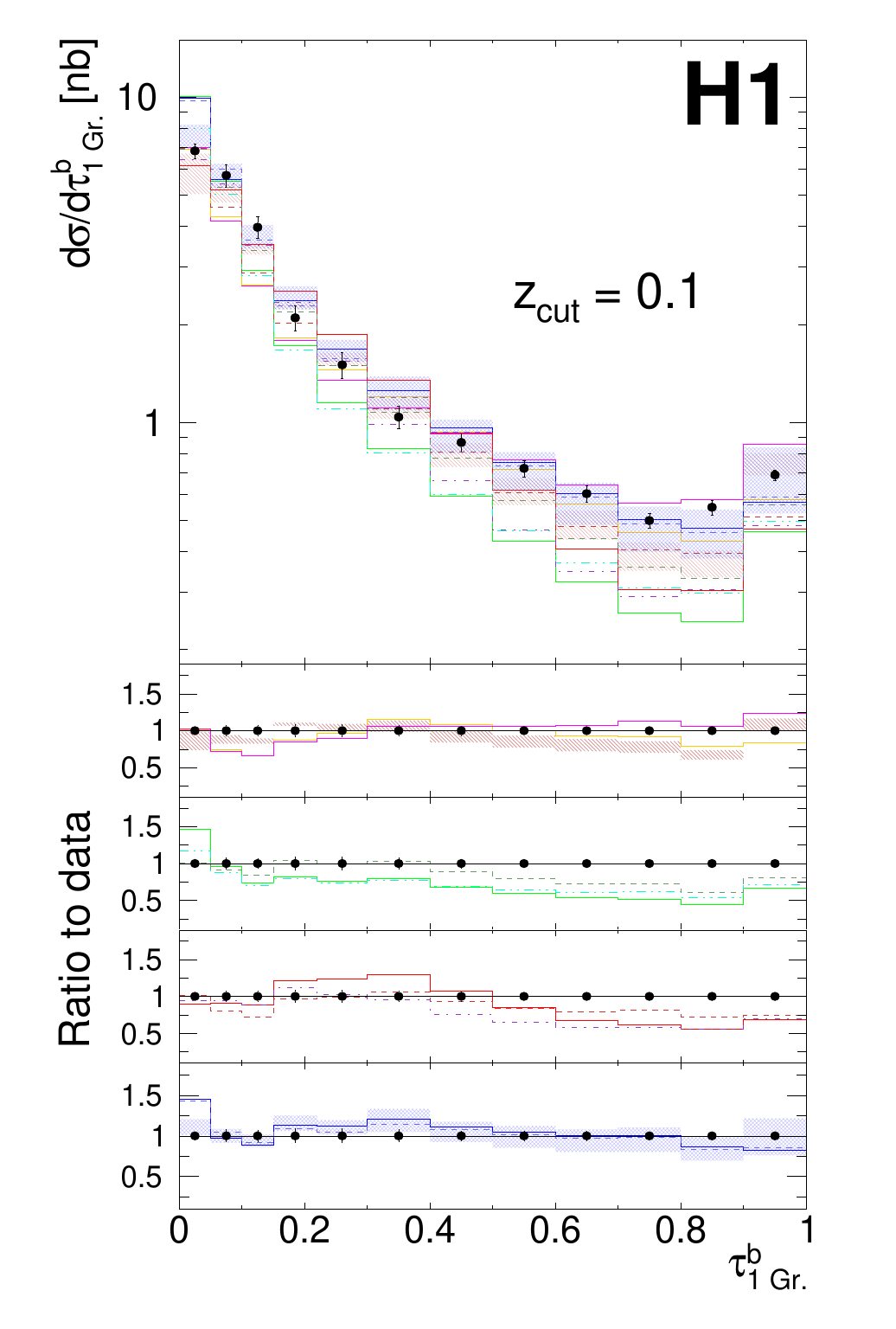}
     \put(60,0){\includegraphics[trim={3.3cm 0 1cm 0},clip,scale=0.4]{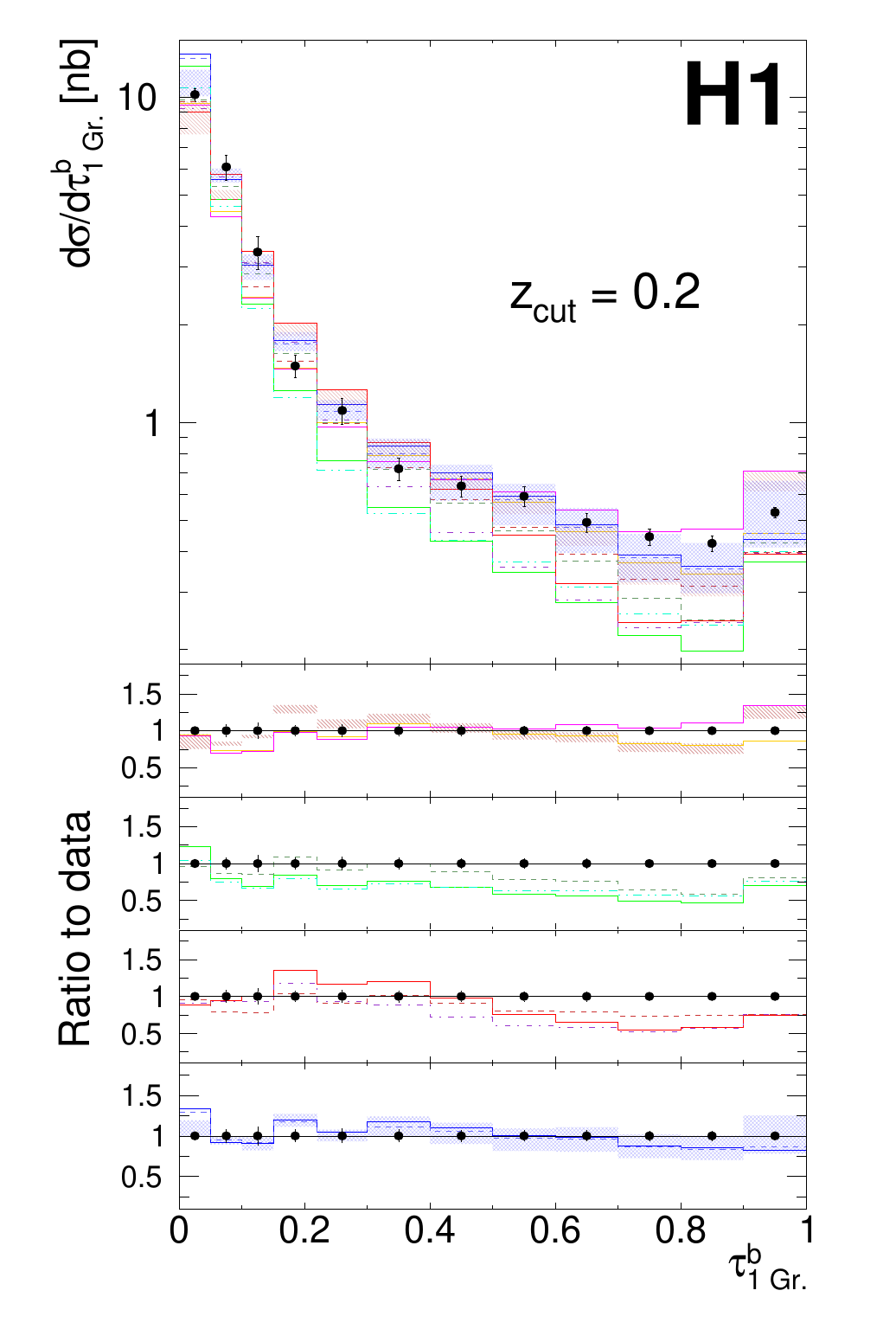}}
     \put(-50,0){\includegraphics[trim={1cm 0 1.1cm 0},clip,scale=0.4]{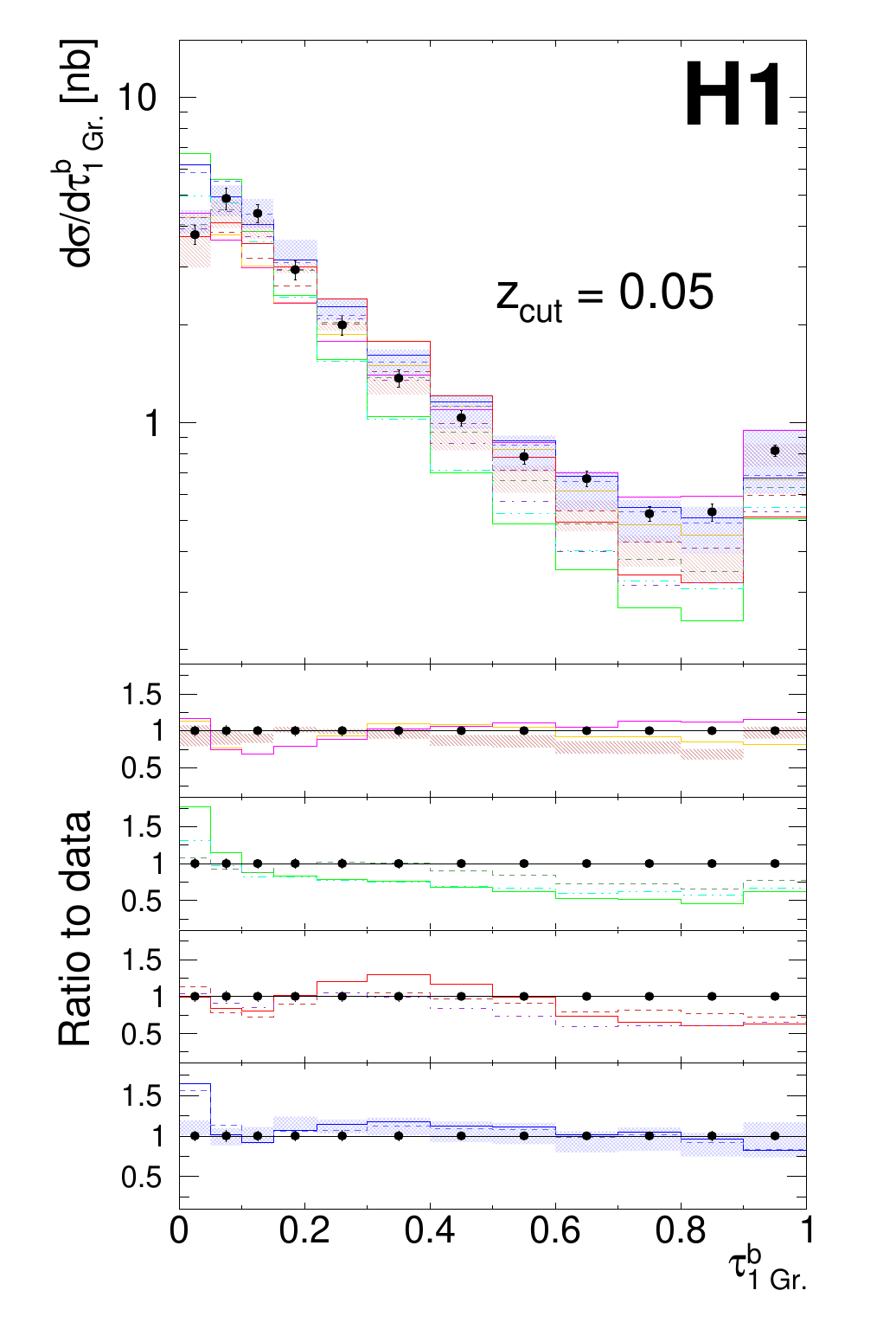}}
  \end{overpic}
\caption{Differential cross section of groomed 1-jettiness \tb in \ep\ DIS at $\sqrts=319~ \mathrm{GeV}$, for $\zcut=0.05$, 0.1 and, 0.2. The phase space is restricted to $\Qsq>150$~\GeVsq~and $0.2 < y < 0.7$. Uncertainty bars on the data show the quadrature sum of the statistical error and systematic uncertainty. The Monte Carlo event generators and pQCD calculations that are compared to the data are described in Section~\ref{sect:Theory}.}
\label{fig:1DGrTaus}
\end{figure*}

Figures~\ref{fig:MC1DGIMs} and~\ref{fig:1DGrTaus} show the single-differential GIM and \tb cross sections, respectively, for $\zcut=0.05$, 0.1, and 0.2. The numerical values of the data points are provided in~\cref{tab:GIM05_1D,tab:GIM1_1D,tab:GIM2_1D,tab:GrTau05_1D,tab:GrTau1_1D,tab:GrTau2_1D}. The GIM distributions exhibit peaks around $\GIMlnone \sim -2$, corresponding to masses of around 20 GeV. The \tb distributions peak at small \tb, around 0.05. These values of GIM and \tb are referred to as the ``peak" region and roughly correspond to events wherein the groomed final state is a single jet. The region $\GIMlnone \gtrsim 1$ and $\tb \gtrsim 0.5$ is referred to henceforth as the ``tail" or ``fixed-order" region and typically corresponds to events with multiple jets or sub-jets that survived grooming. The tail region is sensitive to matrix elements, PDFs, and the color connection between the struck parton and the beam remnant.
The figures show that most of the MC generators underpredict the large mass and large $\tb$ region of the groomed event shape observables. The level of disagreement between the models and the data in this region does not appear to be a strong function of $z_{\mathrm{cut}}$. 

Sherpa 3 better describes the first bin of the groomed \tb distribution compared to Sherpa 2. This could arise from either the improved hadronization model or the addition of intrinsic \kT\ to the initial-state partons. With the improvement of the first bin, Sherpa 3 successfully describes the data within uncertainties across the whole \tb distribution. The 7-point scale variation produces uncertainties around 10\% in the peak region and 30\% in the tail region.

The NNLO+NLL$^\prime$ prediction provides a reasonable description of the single-jet peak region at low \zcut, but the description is poorer at higher \zcut. This may indicate the need for higher resummed accuracy at higher \zcut. The prediction underestimates the cross section in the tail region, where the fixed order calculation is expected to provide an accurate description of the data. 

\subsection{Comparison to SCET predictions}
\label{sect:SingleSCET}

\begin{figure*}[t]
\begin{subfigure}[b]{\textwidth}
    \centering
\includegraphics[width=0.75\textwidth]{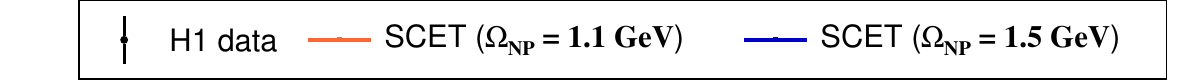}
\end{subfigure}
  \centering
   \begin{overpic}[trim={1.1cm 0 1cm 0},clip,scale=0.35]{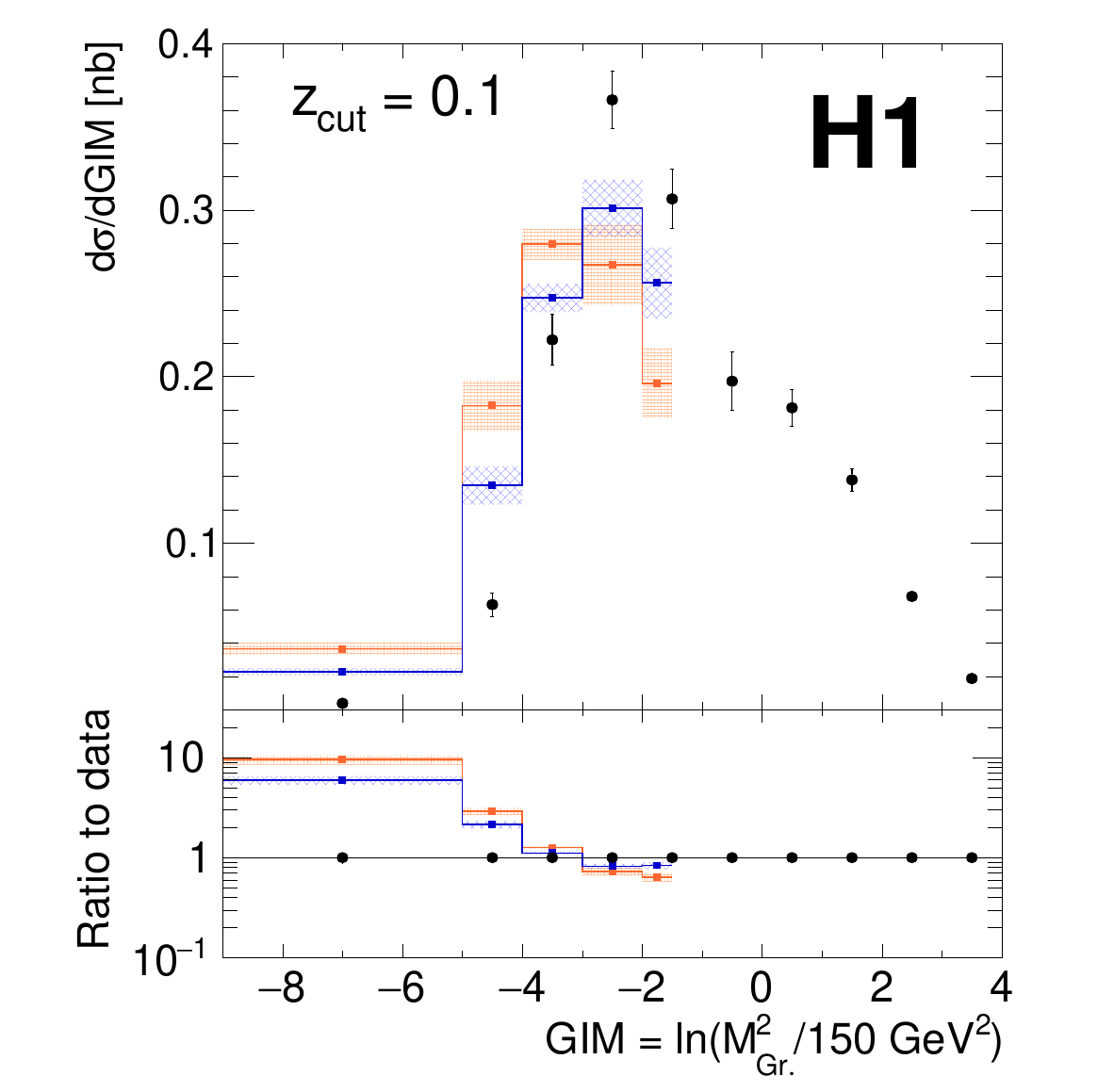}
     \put(89,0){\includegraphics[trim={3.9cm 0 1cm 0},clip,scale=0.35]{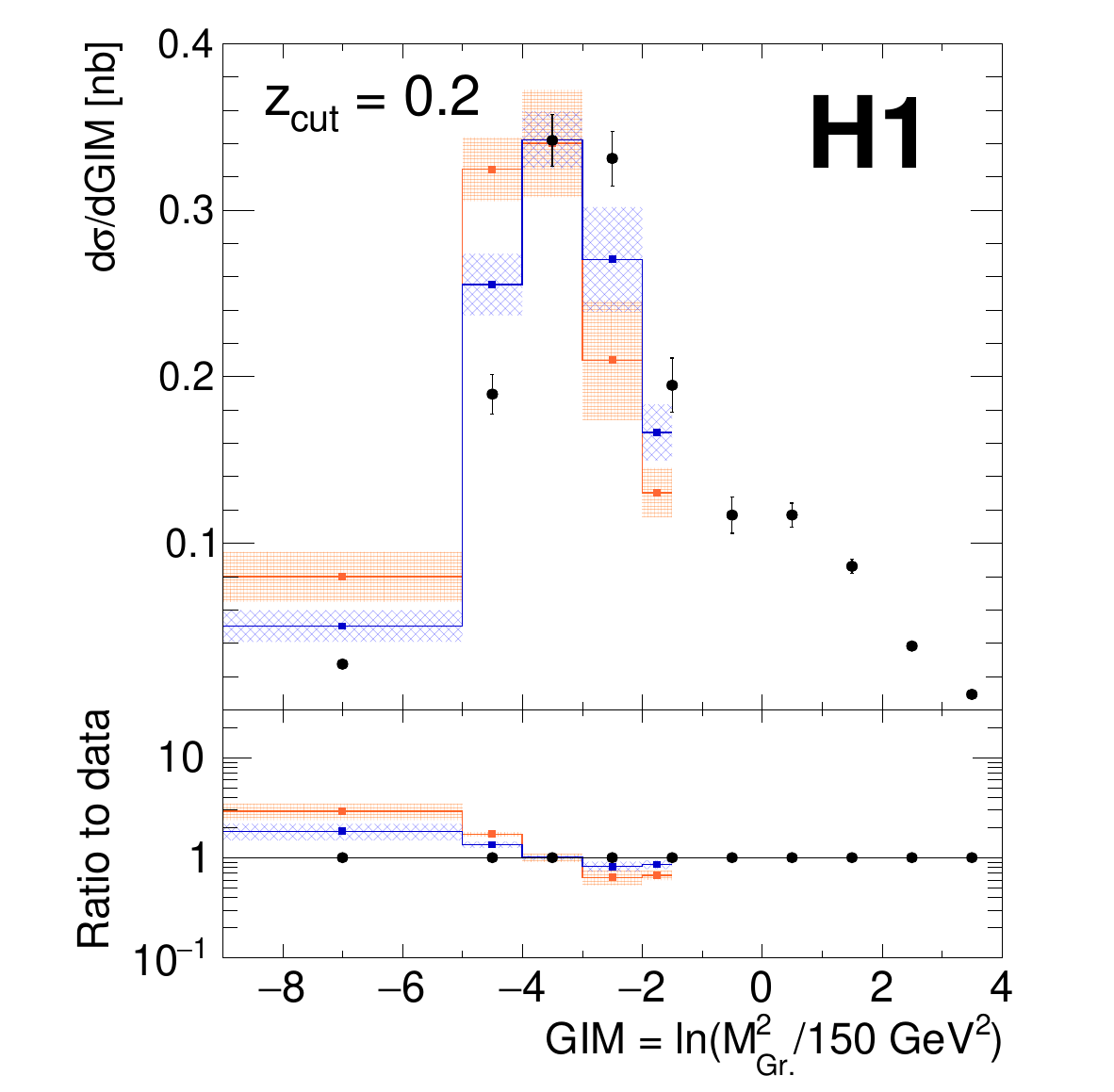}}
     \put(-75,0){\includegraphics[trim={1cm 0 1.5cm 0},clip,scale=0.35]{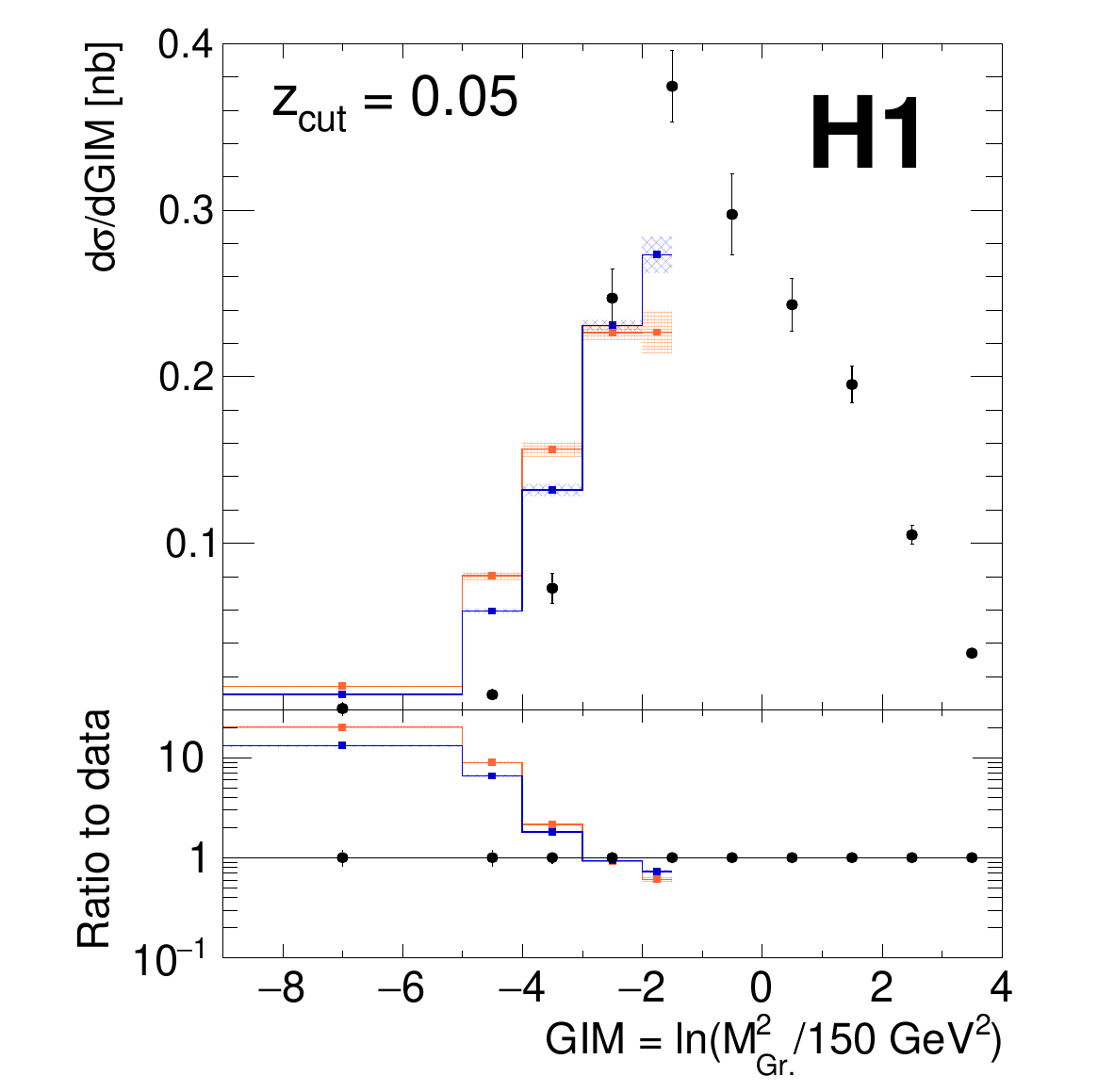}}
  \end{overpic}
\caption{Differential cross section of groomed invariant mass GIM~$=\GIMlnone$ in \ep\ DIS at $\sqrts=319~ \mathrm{GeV}$, for $\zcut=0.05$, 0.1 and, 0.2. The value of \Qminsq is set to 150 \GeVsq. The phase space is restricted to $\Qsq>150$~\GeVsq~and $0.2 < y < 0.7$. Uncertainty bars on the data show the quadrature sum of the statistical error and systematic uncertainty. SCET calculations~\cite{Makris:2021drz} are compared to the data, with value of the shape function mean $\Omega_{\mathrm{NP}}$ of 1.1 GeV and 1.5 GeV. The calculations are normalized to the integral of the data in the region $\mathrm{GIM}<-1$. The shaded bands on the predictions correspond to the associated scale uncertainty.}
\label{fig:SCET1DGIMs}
\end{figure*}

Figure~\ref{fig:SCET1DGIMs} shows the measured GIM single-differential cross section, with SCET calculations in comparison~\cite{Makris:2021drz}. The predictions are normalized to the data in the range $\mathrm{GIM}<-1$ by equating their integrals. Two values of the mean of the non-perturbative shape function, $\Omega_{\mathrm{NP}} = 1.1$ GeV and 1.5 GeV, are used in the prediction. The shape function encapsulates the non-perturbative contribution to the observable resulting from hadronization, which becomes increasingly important at low values of GIM. The prediction has associated scale uncertainties, which are determined by varying all scales in the perturbative prediction by a factor of 2. Note, however, that the uncertainty of the shape function is not evaluated, so that the total theory uncertainty is underestimated at the smaller values of \zcut, where the shape function makes a more significant contribution to the total distribution.

The level of agreement of the calculation with data is limited for $\zcut=0.05$ and 0.1, with better agreement for $\zcut=0.2$. This accords with the expectation that the SCET approximation is valid for $1\gg \zcut \sim \GIM$~\cite{Makris:2021drz}, which is not respected for $\zcut=0.05$ and 0.1. The data likewise prefer $\Omega_\mathrm{NP}=1.5$ GeV, which is expected since the calculation generates on average smaller mass than observed in the data, and the shape function, which accounts for non-perturbative effects, increases the mass relative to the partonic calculation. The value $\Omega_\mathrm{NP}=1.5$ GeV is larger than the naive expectation, $\Omega_\mathrm{NP}\sim1$ GeV~\cite{Makris:2021drz}. The high value of $\Omega_\mathrm{NP}$ may compensate for the effect of gluon jets, which are not included in the calculation.

\begin{figure*}[t]
\begin{center}
\includegraphics[width=16cm]{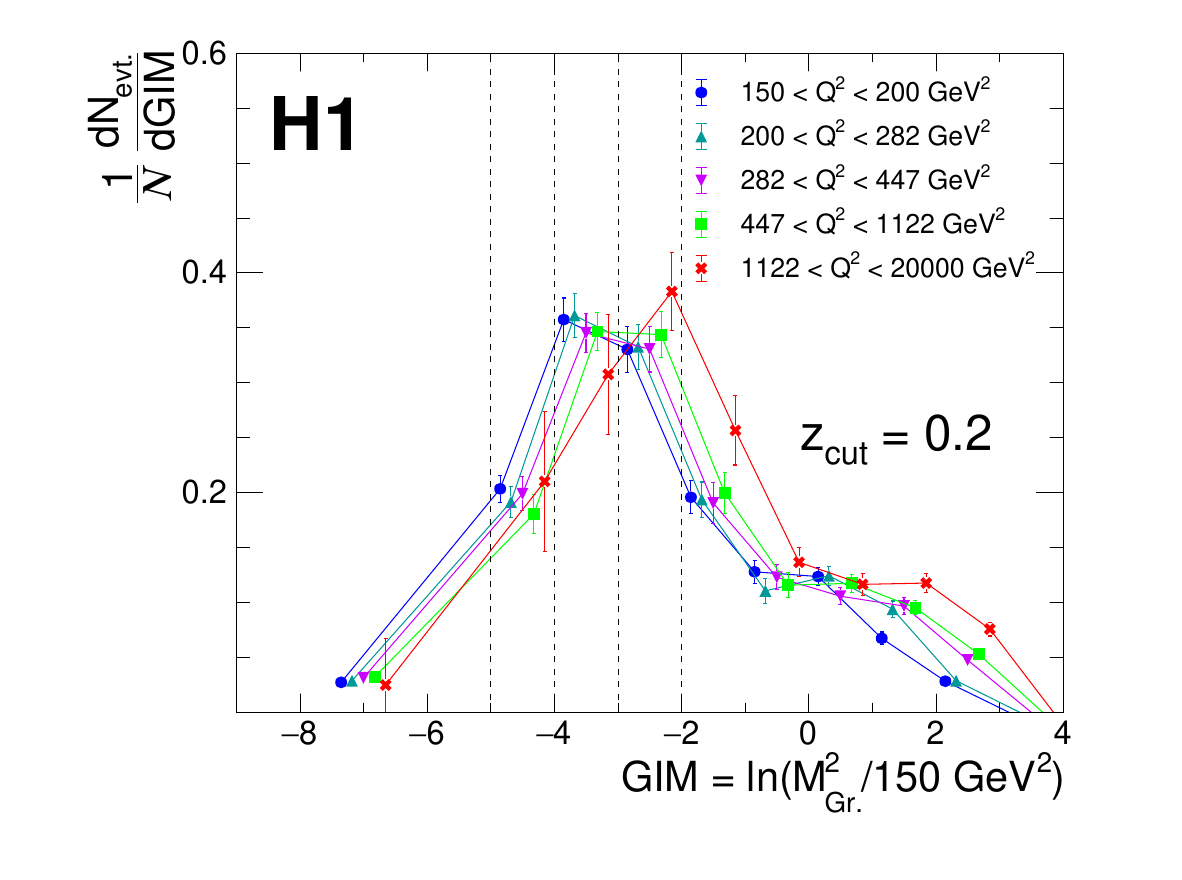}
\caption{Number of events as a function of GIM for \zcut~= 0.2 in several \Qsq\ bins, with integral normalized to unity in the region $\mathrm{GIM}<-2$. \Qminsq Vertical dotted lines show bin boundaries of bins used in the normalization. The data points are offset horizontally for clarity. The bins have been scaled by their width.}
 \label{fig:UniversalityGIM2}
\end{center}
\end{figure*}

The SCET calculation predicts that the shape of the groomed invariant mass distribution is independent of \Qsq\ in the low mass limit, defined by the relation
\begin{equation}
\frac{M^2_{\text{Gr.}}}{Q^2} \ll z_{\mathrm{cut}} \ll 1.
\label{LowMassLimit}
\end{equation}
\noindent
Figure~\ref{fig:UniversalityGIM2}, which tests this prediction, shows the GIM distribution for \zcut~= 0.2 in five bins of \Qsq. The factor \Qminsq is taken to be 150~\GeVsq~for all \Qsq bins. The integrals of all \Qsq distributions are normalized in the region $\mathrm{GIM}<-2$. In this region, the \xB\ and \Qsq-dependence of the cross section occurs only in the component of the event that has been groomed away; the groomed distribution is therefore expected to be invariant with respect to \xB\ and \Qsq. The shape of the distributions shown in Fig.~\ref{fig:UniversalityGIM2} is observed to be independent of \Qsq, in agreement with this prediction.
 

\subsection{Double-differential cross sections}
\label{sect:DoubleSHERPA}

\begin{figure*}[t]
\begin{center}
\includegraphics[width=12cm]{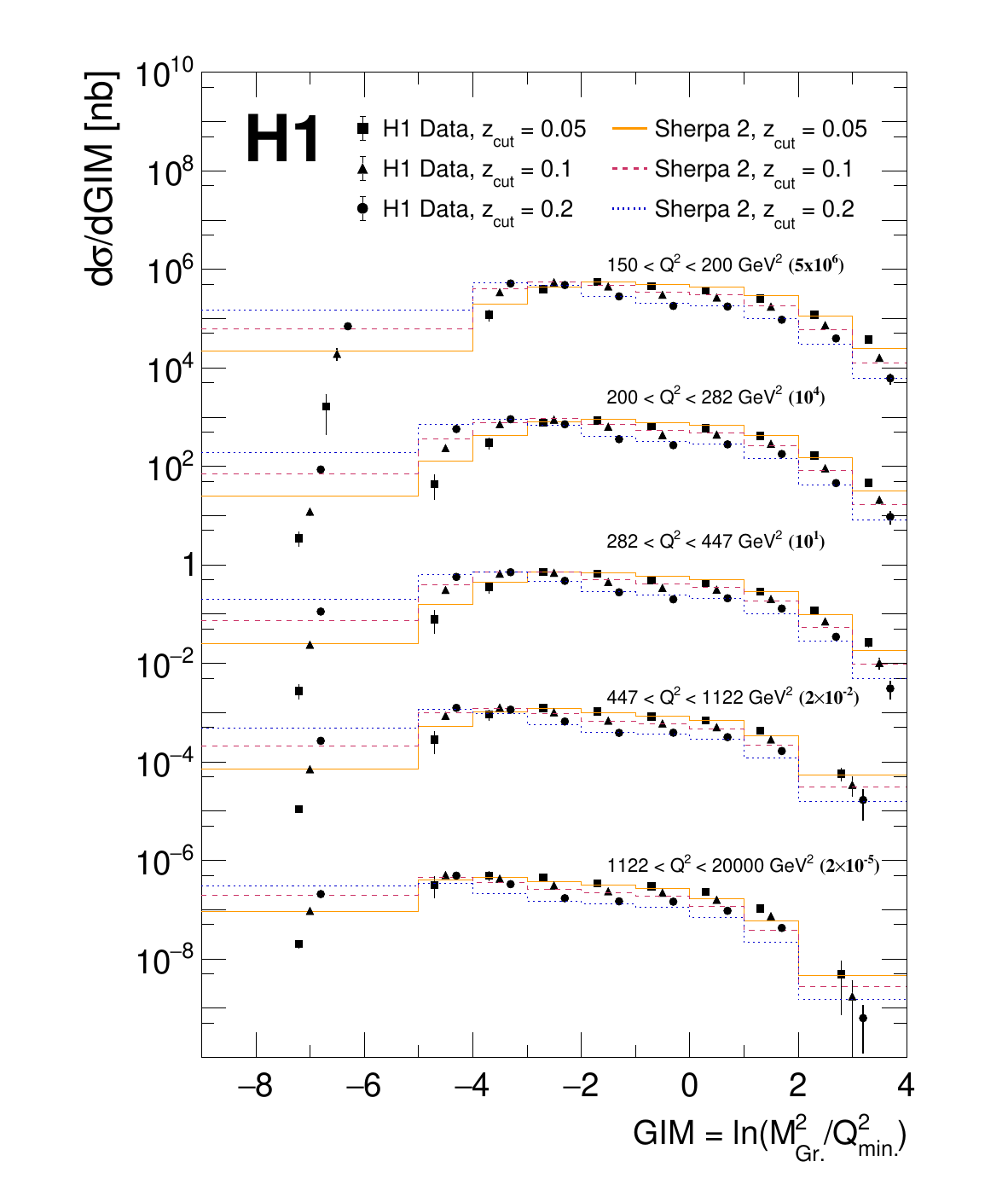}
\caption{Double-differential groomed invariant mass distributions for $\zcut = 0.05$, 0.1, and 0.2, for five bins in \Qsq. The value of \Qminsq is taken to be the lowest value of \Qsq in the bin, e.g. 1122~\GeVsq~for the highest \Qsq bin. Uncertainty bars on the data show the quadrature sum of the statistical and systematic uncertainties. The data points are horizontally offset from the bin center for visibility. The lines represent predictions from Sherpa 2 using the cluster hadronization model. The five distributions at different values of \Qsq are vertically offset by the factor given in parentheses.}
 \label{fig:2DGIM}
\end{center}
\end{figure*}

\begin{figure*}[t]
\begin{center}
\includegraphics[width=12cm]{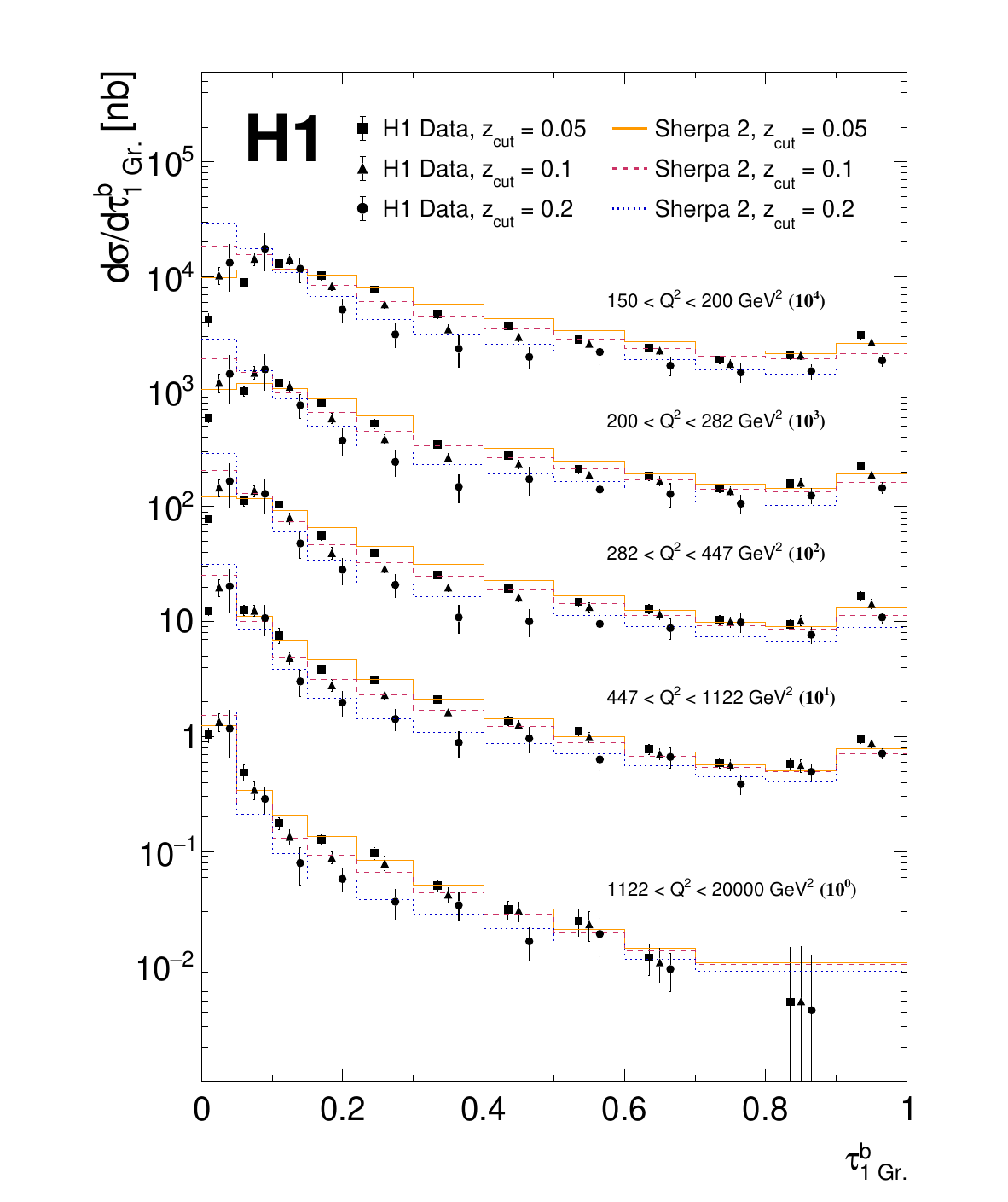}
\caption{Double-differential \tb cross section for $\zcut = 0.05$, 0.1, and 0.2, presented in five bins in \Qsq. Further details are given in the caption of Fig.~\ref{fig:2DGIM}}
 \label{fig:2DGrTau}
\end{center}
\end{figure*}

Figures~\ref{fig:2DGIM} and \ref{fig:2DGrTau} show double-differential cross sections of GIM and \tb, alongside calculations from Sherpa 2 with the cluster hadronization model for comparison. The data are presented in five bins of $Q^2$ and at three values of \zcut. The binning used in these figures is presented in Tab.~\ref{tab:binnings}.
\begin{table*}[t]
  \footnotesize
  \begin{center}
  
    \begin{tabular}{lc}
      \hline
      Observable & Binning \\
      \hline
    Standard \tb   &  $[0.0, 0.05, 0.10, 0.15, 0.22, 0.3, 0.4, 0.5, 0.6, 0.7, 0.8, 0.9, 1.00]$ \\ 
    Reduced \tb   &  $[0.0, 0.05, 0.10, 0.15, 0.22, 0.3, 0.4, 0.5, 0.6, 0.7, 1.00]$ \\
    Standard \GIMln  &  $[-9,-5,-4,-3,-2,-1,0,1,2,3,4]$  \\ 
    Reduced Low \GIMln  &  $[-9,-4,-3,-2,-1,0,1,2,3,4]$\\
    Reduced High \GIMln  &  $[-9,-5,-4,-3,-2,-1,0,1,2,4]$\\
    $\Qsq$&  $[150, 200, 282, 447, 1122, 20000]$ \GeVsq \\
      \hline
    \end{tabular}   
        \caption{
      Binnings used in the analysis. The reduced binnings are used only in the low-statistics regions of the double-differential result.
    }
    \label{tab:binnings}
    \end{center}
\end{table*}
Bins with very small event counts were merged with the neighboring bins. The numerical values of the data are provided in~\cref{tab:GIM2D_Q1,tab:GIM2D_Q2,tab:GIM2D_Q3,tab:GIM2D_Q4,tab:GIM2D_Q5,tab:GrTau2D_Q1,tab:GrTau2D_Q2,tab:GrTau2D_Q3,tab:GrTau2D_Q4,tab:GrTau2D_Q5}.

A reasonable agreement between the predictions of Sherpa 2 and the data is found in the majority of bins, with some tension observed at very low \tb and GIM. The description of the peak region of the \tb distribution can be seen to improve with \Qsq. 

At higher values of \Qsq, the mean values of the event shape distributions decrease, in accordance with the expectation from QCD. These measurements may provide new constraints on the strong coupling constant \as. 


\section{Summary}
\label{sect:Summary}

This paper presents the first measurement of groomed event shape observables in deep inelastic $ep$ collisions. Measurements of the invariant mass and 1-jettiness \tb of the groomed hadronic final state are reported for \emp\ collisions at $\sqrts=319$ GeV, for events selected with $\Qsq > 150$ GeV$^2$ and $ 0.2 < y < 0.7$. Events are clustered using the Centauro jet algorithm, and results are presented for values of the grooming parameter $\zcut=0.05$, 0.1, and 0.2. Cross sections are reported single- and double-differentially. 

Event grooming suppresses non-perturbative contributions to event shape distributions in a theoretically well-controlled way. Comparisons of Monte Carlo models and analytical pQCD calculations to these data therefore provide significant new tests of their implementation of both perturbative and non-perturbative processes.

Two of the models that are commonly compared to HERA data, Rapgap and Djangoh, can describe the data in the fixed-order tail regions of the groomed event shape distributions but underestimate the single-jet peak regions. More recently developed models, Pythia 8, Herwig 7, and Sherpa 2, underestimate the fixed-order tail region. The agreement of the models with the data in the low-mass or low-\tb region improves for higher \zcut. Sherpa 3 accurately describes the full distribution of the groomed 1-jettiness within the experimental and theoretical uncertainties. 


The numerical predictions of the SCET calculation fail for the lower \zcut\ values. Comparison of the calculation with data indicates overall a preference for a larger value of the non-perturbative shape function $\Omega_{\mathrm{NP}}$, suggesting that hadronization and other non-perturbative effects are significant in the single-jet limit. The prediction of Ref.~\cite{Makris:2021drz}, that the shape of the low-mass region is independent of the hard scale $Q^2$, is found to hold in spite of the disagreement between the numerical predictions and the data.

Event shapes are sensitive to many aspects of QCD final states, making them valuable input for tuning MC event generators. However, such models contain many parameters and determining their optimum values is challenging, since a given effect can often be correctly described by tuning multiple different parameters. The introduction of grooming suppresses the contribution of certain event components, such as the proton beam fragmentation, and causes others, such as the contribution from soft radiation, to scale with \zcut.


An accurate description of DIS in MC event generators is crucial for the scientific program of the upcoming Electron-Ion Collider~\cite{AbdulKhalek:2021gbh}. Future facilities currently under discussion, the LHeC~\cite{LHeCStudyGroup:2012zhm,LHeC:2020van} and FCC-eh~\cite{FCC:2018byv}, likewise will require precise MC modeling of the DIS hadronic final state to achieve their physics goals. The groomed event shape distributions reported here provide new, differential constraints for the tuning of MC models and offer the possibility for extracting PDFs and fundamental QCD parameters such as $\alpha_s$.

\section*{Acknowledgements}
We are grateful to the HERA machine group whose outstanding efforts
have made this experiment possible. We thank the engineers and
technicians for their work in constructing and maintaining the H1
detector, our funding agencies for financial support, the DESY
technical staff for continual assistance and the DESY directorate for
support and for the hospitality which they extend to the non-DESY members of the collaboration.
We express our thanks to all those involved in
securing not only the H1 data but also the software and working
environment for long term use,
allowing the unique H1 data set to continue to
be explored. The transfer from experiment specific to central resources with long term support,
including both storage and batch systems, has also been crucial to this enterprise.
We therefore also acknowledge the role played by DESY-IT and all
people involved during this transition and their
future role in the years to come.

We would like to thank
Silvia Ferrario Ravasio,
Christopher Lee,
Yiannis Makris,
Simon Plätzer,
Christian Preuss,
and
Felix Ringer
for
many valuable comments and discussions, for providing us with theoretical
predictions, or for help with the predictions.

\par $^{f1}$ supported by the U.S. DOE Office of Science
\par $^{f2}$ supported by FNRS-FWO-Vlaanderen, IISN-IIKW and IWT and by Interuniversity Attraction Poles Programme, Belgian Science Policy
\par $^{f3}$ supported by the UK Science and Technology Facilities Council, and formerly by the UK Particle Physics and Astronomy Research Council
\par $^{f4}$ supported by the Romanian National Authority for Scientific Research under the contract PN 09370101
\par $^{f5}$ supported by the Bundesministerium für Bildung und Forschung, FRG, under contract numbers 05H09GUF, 05H09VHC, 05H09VHF, 05H16PEA
\par $^{f6}$ partially supported by Polish Ministry of Science and Higher Education, grant DPN/N168/DESY/2009
\par $^{f7}$ partially supported by Ministry of Science of Montenegro, no. 05-1/3-3352
\par $^{f8}$ supported by the Ministry of Education of the Czech Republic under the project INGO-LG14033
\par $^{f9}$ supported by CONACYT, México, grant 48778-F
\par $^{f10}$ supported by the Swiss National Science Foundation



\bibliography{desy24-036}

\FloatBarrier

\section{Tables}
\label{sect:Tables}
Numerical data are provided for the single-differential cross sections as a function of the groomed invariant mass for grooming parameter $\zcut = 0.05$ in Tab.~\ref{tab:GIM05_1D}, for $\zcut = 0.1$ in Tab.~\ref{tab:GIM1_1D}, and for $\zcut = 0.2$ in Tab.~\ref{tab:GIM2_1D}. Similarly, numerical cross-section data are presented as a function of the groomed 1-jettiness for grooming parameter $\zcut = 0.05$ in Tab.~\ref{tab:GrTau05_1D}, for $\zcut = 0.1$ in Tab.~\ref{tab:GrTau1_1D}, and for $\zcut = 0.2$ in Tab.~\ref{tab:GrTau2_1D}.


\begin{table*}[tbhp]
\tiny
\centering
\begin{tabular}{cc|c|cccccccccc|c}
\toprule
\multicolumn{13}{c}{1D Groomed invariant mass cross section for $\zcut = 0.05$}\\
\midrule
\multicolumn{2}{c|}{ln($M^2_{Gr.}/150~\GeVsq$) range} & Results & \multicolumn{10}{c|}{Uncertainty} & QED Factor\\
\cmidrule(lr){1-2}\cmidrule(lr){3-3}\cmidrule(lr){4-13}\cmidrule{14-14}
$\mathrm{Bin~Min.}$ & $\mathrm{Bin~Max.}$ & $d\sigma/d\mathrm{GIM}$ & Total & Stat. & JES & RCES & $\theta_{\mathrm{HFS}}$ & $E_e$ & $\theta_e$ & Lumi & Reg. & Model & $c_\mathrm{QED}$\\
& & $[\text{pb}]$ & $_{[\%]}$ & $_{[\%]}$ & $_{[\%]}$ & $_{[\%]}$ & $_{[\%]}$ & $_{[\%]}$ & $_{[\%]}$ & $_{[\%]}$ &  $_{[\%]}$ & $_{[\%]}$ &\\
\midrule
-9 & -5 & 7.09e-01 & 20.8 & 16.8 & 1.6 & 0.6 & -0.0 & -0.5 & -0.1 & 2.7 & 2.3 & 9.1 & 1.184\\
-5 & -4 & 9.02e+00 & 19.8 & 9.4 & 3.6 & 0.6 & -0.1 & -1.0 & 0.2 & 2.7 & 2.2 & 16.2 & 1.158\\
-4 & -3 & 7.30e+01 & 13.1 & 3.1 & 3.6 & 0.7 & 0.0 & -0.8 & 0.2 & 2.7 & 1.6 & 10.9 & 1.163\\
-3 & -2 & 2.47e+02 & 8.1 & 1.5 & 1.0 & 0.7 & 0.1 & -0.7 & 0.2 & 2.7 & 2.4 & 5.6 & 1.160\\
-2 & -1 & 3.74e+02 & 7.1 & 1.2 & -0.5 & 0.5 & 0.1 & -0.8 & 0.3 & 2.7 & 2.5 & 4.1 & 1.162\\
-1 & 0 & 2.97e+02 & 8.8 & 0.9 & 1.0 & 0.3 & 0.2 & -1.4 & 0.4 & 2.7 & 1.6 & 7.2 & 1.156\\
0 & 1 & 2.43e+02 & 7.4 & 0.9 & 1.9 & -0.7 & 0.0 & -1.7 & 0.4 & 2.7 & 1.4 & 5.0 & 1.158\\
1 & 2 & 1.95e+02 & 6.5 & 1.2 & 1.4 & -1.3 & -0.1 & -1.1 & 0.2 & 2.7 & 0.9 & 4.0 & 1.157\\
2 & 3 & 1.05e+02 & 6.1 & 1.6 & 0.1 & -0.7 & -0.3 & -0.7 & 0.2 & 2.7 & 0.5 & 4.0 & 1.161\\
3 & 4 & 3.39e+01 & 6.2 & 3.5 & -1.3 & -0.3 & -0.7 & -0.3 & 0.1 & 2.7 & 0.4 & 2.4 & 1.171\\

\bottomrule
\end{tabular}
\caption{Single-differential groomed invariant mass cross sections for $\zcut = 0.05$. The statistical error represents the outcome of the replica method described in Sec.~\ref{sect:Uncertainties}. The sources of systematic uncertainty are described in detail in Sec.~\ref{sect:Uncertainties}. The total uncertainty on the data is the quadrature sum of the statistical and systematic errors. $c_{\mathrm{QED}}$ is the QED correction factor as derived from HERACLES; the radiative cross section can be recovered by dividing the given data by the corresponding value of $c_{\mathrm{QED}}$.}
\label{tab:GIM05_1D} 
\end{table*}

\begin{table*}[tbhp]
\tiny
\centering
\begin{tabular}{cc|c|cccccccccc|c}
\toprule
\multicolumn{13}{c}{1D Groomed invariant mass cross section for $\zcut = 0.1$}\\
\midrule
\multicolumn{2}{c|}{ln($M^2_{Gr.}/150~\GeVsq$) range} & Results & \multicolumn{10}{c|}{Uncertainty} & QED Factor\\
\cmidrule(lr){1-2}\cmidrule(lr){3-3}\cmidrule(lr){4-13}\cmidrule{14-14}
$\mathrm{Bin~Min.}$ & $\mathrm{Bin~Max.}$ & $d\sigma/d\mathrm{GIM}$ & Total & Stat. & JES & RCES & $\theta_{\mathrm{HFS}}$ & $E_e$ & $\theta_e$ & Lumi & Reg. & Model & $c_\mathrm{QED}$\\
& & $[\text{pb}]$ & $_{[\%]}$ & $_{[\%]}$ & $_{[\%]}$ & $_{[\%]}$ & $_{[\%]}$ & $_{[\%]}$ & $_{[\%]}$ & $_{[\%]}$ &  $_{[\%]}$ & $_{[\%]}$ &\\
\midrule
-9 & -5 & 3.83e+00 & 13.6 & 6.8 & 2.6 & 0.3 & -0.0 & -0.9 & 0.1 & 2.7 & 1.6 & 9.9 & 1.167\\
-5 & -4 & 6.31e+01 & 11.8 & 4.3 & 3.5 & 0.3 & -0.0 & -0.9 & 0.2 & 2.7 & 2.0 & 9.0 & 1.161\\
-4 & -3 & 2.22e+02 & 8.0 & 1.9 & 2.5 & 0.3 & -0.0 & -0.9 & 0.2 & 2.7 & 1.9 & 5.0 & 1.161\\
-3 & -2 & 3.66e+02 & 6.1 & 1.4 & -0.0 & 0.3 & 0.1 & -0.7 & 0.3 & 2.7 & 1.3 & 3.2 & 1.160\\
-2 & -1 & 3.07e+02 & 6.8 & 1.3 & -0.6 & 0.4 & 0.1 & -1.0 & 0.3 & 2.7 & 1.0 & 4.7 & 1.161\\
-1 & 0 & 1.97e+02 & 9.6 & 1.1 & 1.5 & 0.0 & 0.1 & -1.7 & 0.4 & 2.7 & 1.1 & 8.1 & 1.157\\
0 & 1 & 1.81e+02 & 7.1 & 1.2 & 2.0 & -0.9 & 0.1 & -1.7 & 0.4 & 2.7 & 0.8 & 4.6 & 1.158\\
1 & 2 & 1.38e+02 & 6.0 & 1.5 & 0.9 & -0.9 & -0.0 & -0.9 & 0.2 & 2.7 & 0.5 & 3.6 & 1.157\\
2 & 3 & 6.80e+01 & 5.8 & 2.0 & -0.5 & -0.3 & -0.2 & -0.4 & 0.1 & 2.7 & 0.4 & 3.3 & 1.163\\
3 & 4 & 1.87e+01 & 6.3 & 4.2 & -1.5 & -0.3 & -0.6 & -0.2 & 0.1 & 2.7 & 0.3 & 2.3 & 1.170\\

\bottomrule
\end{tabular}
\caption{Same as Table~\ref{tab:GIM05_1D} for the GIM cross section at $\zcut = 0.1$.}
\label{tab:GIM1_1D} 
\end{table*}

\begin{table*}[tbhp]
\tiny
\centering
\begin{tabular}{cc|c|cccccccccc|c}
\toprule
\multicolumn{13}{c}{1D Groomed invariant mass cross section for $\zcut = 0.2$}\\
\midrule
\multicolumn{2}{c|}{ln($M^2_{Gr.}/150~\GeVsq$) range} & Results & \multicolumn{10}{c|}{Uncertainty} & QED Factor\\
\cmidrule(lr){1-2}\cmidrule(lr){3-3}\cmidrule(lr){4-13}\cmidrule{14-14}
$\mathrm{Bin~Min.}$ & $\mathrm{Bin~Max.}$ & $d\sigma/d\mathrm{GIM}$ & Total & Stat. & JES & RCES & $\theta_{\mathrm{HFS}}$ & $E_e$ & $\theta_e$ & Lumi & Reg. & Model & $c_\mathrm{QED}$\\
& & $[\text{pb}]$ & $_{[\%]}$ & $_{[\%]}$ & $_{[\%]}$ & $_{[\%]}$ & $_{[\%]}$ & $_{[\%]}$ & $_{[\%]}$ & $_{[\%]}$ &  $_{[\%]}$ & $_{[\%]}$ &\\
\midrule
-9 & -5 & 2.74e+01 & 7.9 & 3.7 & 1.8 & 0.2 & -0.0 & -1.1 & 0.2 & 2.7 & 0.6 & 4.2 & 1.166\\
-5 & -4 & 1.89e+02 & 7.5 & 2.7 & 2.4 & 0.2 & -0.0 & -1.0 & 0.2 & 2.7 & 0.3 & 4.2 & 1.160\\
-4 & -3 & 3.42e+02 & 6.1 & 1.5 & 1.1 & 0.1 & -0.0 & -1.0 & 0.3 & 2.7 & 0.9 & 2.9 & 1.161\\
-3 & -2 & 3.31e+02 & 6.2 & 1.4 & -0.7 & 0.0 & 0.1 & -0.9 & 0.3 & 2.7 & 1.3 & 3.5 & 1.159\\
-2 & -1 & 1.95e+02 & 9.0 & 1.5 & -0.5 & 0.4 & 0.1 & -1.2 & 0.3 & 2.7 & 1.0 & 7.6 & 1.157\\
-1 & 0 & 1.17e+02 & 9.9 & 1.4 & 1.9 & -0.1 & 0.1 & -1.9 & 0.4 & 2.7 & 0.7 & 8.4 & 1.156\\
0 & 1 & 1.17e+02 & 7.0 & 1.6 & 1.7 & -0.8 & 0.1 & -1.7 & 0.4 & 2.7 & 0.5 & 4.4 & 1.159\\
1 & 2 & 8.61e+01 & 6.1 & 1.9 & 0.1 & -0.7 & -0.0 & -0.6 & 0.1 & 2.7 & 0.4 & 3.8 & 1.158\\
2 & 3 & 3.82e+01 & 6.1 & 2.6 & -0.7 & -0.3 & -0.1 & -0.2 & 0.1 & 2.7 & 0.3 & 3.4 & 1.165\\
3 & 4 & 9.20e+00 & 7.6 & 5.5 & -1.6 & -0.1 & -0.4 & -0.1 & 0.1 & 2.7 & 0.2 & 2.8 & 1.174\\

\bottomrule
\end{tabular}
\caption{Same as Table~\ref{tab:GIM05_1D} for the GIM cross section at $\zcut = 0.2$.}
\label{tab:GIM2_1D} 
\end{table*}

\begin{table*}[tbhp]
\tiny
\centering
\begin{tabular}{cc|c|cccccccccc|c}
\toprule
\multicolumn{13}{c}{1D $\tb$ cross section for $\zcut = 0.05$}\\
\midrule
\multicolumn{2}{c|}{$\tb$ range} & Results & \multicolumn{10}{c|}{Uncertainty} & QED Factor\\
\cmidrule(lr){1-2}\cmidrule(lr){3-3}\cmidrule(lr){4-13}\cmidrule{14-14}
$\mathrm{Bin~Min.}$ & $\mathrm{Bin~Max.}$ & $d\sigma/d\mathrm{\tb}$ & Total & Stat. & JES & RCES & $\theta_{\mathrm{HFS}}$ & $E_e$ & $\theta_e$ & Lumi & Reg. & Model & $c_\mathrm{QED}$\\
& & $[\text{pb}]$ & $_{[\%]}$ & $_{[\%]}$ & $_{[\%]}$ & $_{[\%]}$ & $_{[\%]}$ & $_{[\%]}$ & $_{[\%]}$ & $_{[\%]}$ &  $_{[\%]}$ & $_{[\%]}$ &\\
\midrule
0.00 & 0.05 & 3.77e+03 & 7.9 & 0.9 & 2.0 & 0.3 & 0.0 & -1.1 & 0.1 & 2.7 & 3.8 & 4.0 & 1.181\\
0.05 & 0.10 & 4.88e+03 & 8.4 & 0.9 & 0.8 & 0.7 & 0.1 & -0.9 & 0.3 & 2.7 & 6.0 & 3.5 & 1.156\\
0.10 & 0.15 & 4.39e+03 & 7.4 & 1.1 & 0.2 & 0.4 & 0.1 & -0.8 & 0.3 & 2.7 & 3.8 & 4.0 & 1.157\\
0.15 & 0.22 & 2.94e+03 & 7.6 & 1.8 & 0.8 & -0.2 & 0.1 & -1.0 & 0.3 & 2.7 & 5.0 & 2.8 & 1.157\\
0.22 & 0.30 & 1.99e+03 & 7.9 & 2.3 & 1.4 & -0.2 & -0.0 & -1.2 & 0.4 & 2.7 & 4.9 & 2.9 & 1.157\\
0.30 & 0.40 & 1.37e+03 & 7.2 & 2.2 & 1.5 & -0.4 & 0.0 & -1.5 & 0.4 & 2.7 & 3.7 & 3.1 & 1.155\\
0.40 & 0.50 & 1.03e+03 & 6.3 & 2.8 & 1.3 & -0.6 & -0.1 & -1.6 & 0.4 & 2.7 & 2.5 & 1.6 & 1.153\\
0.50 & 0.60 & 7.84e+02 & 6.3 & 3.3 & 1.2 & -0.5 & 0.0 & -1.6 & 0.4 & 2.7 & 1.7 & 1.2 & 1.156\\
0.60 & 0.70 & 6.70e+02 & 6.3 & 3.8 & 0.7 & -0.6 & -0.1 & -1.4 & 0.3 & 2.7 & 1.0 & 1.5 & 1.155\\
0.70 & 0.80 & 5.23e+02 & 6.5 & 4.0 & 0.2 & -0.4 & 0.1 & -1.1 & 0.4 & 2.7 & 0.7 & 1.4 & 1.156\\
0.80 & 0.90 & 5.30e+02 & 6.9 & 3.5 & 0.8 & -1.2 & -0.1 & -0.7 & 0.3 & 2.7 & 0.3 & 3.6 & 1.155\\
0.90 & 1.00 & 8.18e+02 & 5.3 & 1.5 & -0.7 & -0.1 & 0.1 & 0.4 & 0.1 & 2.7 & 0.2 & 1.9 & 1.157\\

\bottomrule
\end{tabular}
\caption{Same as Table~\ref{tab:GIM05_1D} for the \tb cross section at $\zcut = 0.05$.}
\label{tab:GrTau05_1D} 
\end{table*}

\begin{table*}[tbhp]
\tiny
\centering
\begin{tabular}{cc|c|cccccccccc|c}
\toprule
\multicolumn{13}{c}{1D $\tb$ cross section for $\zcut = 0.1$}\\
\midrule
\multicolumn{2}{c|}{$\tb$ range} & Results & \multicolumn{10}{c|}{Uncertainty} & QED Factor\\
\cmidrule(lr){1-2}\cmidrule(lr){3-3}\cmidrule(lr){4-13}\cmidrule{14-14}
$\mathrm{Bin~Min.}$ & $\mathrm{Bin~Max.}$ & $d\sigma/d\mathrm{\tb}$ & Total & Stat. & JES & RCES & $\theta_{\mathrm{HFS}}$ & $E_e$ & $\theta_e$ & Lumi & Reg. & Model & $c_\mathrm{QED}$\\
& & $[\text{pb}]$ & $_{[\%]}$ & $_{[\%]}$ & $_{[\%]}$ & $_{[\%]}$ & $_{[\%]}$ & $_{[\%]}$ & $_{[\%]}$ & $_{[\%]}$ &  $_{[\%]}$ & $_{[\%]}$ &\\
\midrule
0.00 & 0.05 & 6.83e+03 & 6.8 & 0.6 & 1.9 & 0.1 & -0.0 & -1.0 & 0.2 & 2.7 & 3.2 & 2.6 & 1.172\\
0.05 & 0.10 & 5.74e+03 & 8.6 & 0.8 & 0.6 & 0.4 & 0.1 & -0.8 & 0.3 & 2.7 & 6.5 & 3.1 & 1.156\\
0.10 & 0.15 & 3.98e+03 & 8.4 & 1.4 & -0.4 & 0.5 & 0.1 & -0.8 & 0.4 & 2.7 & 4.9 & 4.8 & 1.155\\
0.15 & 0.22 & 2.09e+03 & 9.6 & 2.4 & 0.8 & 0.2 & 0.1 & -1.4 & 0.3 & 2.7 & 7.0 & 3.7 & 1.158\\
0.22 & 0.30 & 1.50e+03 & 9.9 & 3.0 & 1.5 & -0.3 & -0.0 & -1.4 & 0.4 & 2.7 & 7.0 & 3.9 & 1.158\\
0.30 & 0.40 & 1.04e+03 & 8.5 & 3.0 & 1.8 & -0.4 & 0.0 & -1.6 & 0.3 & 2.7 & 5.3 & 3.1 & 1.155\\
0.40 & 0.50 & 8.66e+02 & 7.2 & 3.5 & 1.3 & -0.6 & 0.0 & -1.6 & 0.5 & 2.7 & 3.5 & 1.9 & 1.153\\
0.50 & 0.60 & 7.21e+02 & 7.0 & 4.1 & 1.2 & -0.4 & 0.0 & -1.6 & 0.4 & 2.7 & 2.3 & 1.2 & 1.156\\
0.60 & 0.70 & 6.03e+02 & 6.8 & 4.6 & 0.6 & -0.5 & -0.1 & -1.3 & 0.3 & 2.7 & 1.3 & 1.2 & 1.155\\
0.70 & 0.80 & 4.98e+02 & 6.7 & 4.2 & 0.1 & -0.3 & 0.1 & -1.0 & 0.4 & 2.7 & 0.8 & 1.2 & 1.155\\
0.80 & 0.90 & 5.47e+02 & 6.3 & 3.5 & 0.8 & -1.2 & -0.1 & -0.5 & 0.2 & 2.7 & 0.4 & 2.2 & 1.156\\
0.90 & 1.00 & 6.88e+02 & 5.0 & 1.6 & -0.8 & -0.0 & 0.1 & 0.4 & 0.1 & 2.7 & 0.3 & 1.4 & 1.156\\
\bottomrule
\end{tabular}
\caption{Same as Table~\ref{tab:GIM05_1D} for the \tb cross section at $\zcut = 0.1$.}
\label{tab:GrTau1_1D} 
\end{table*}

\begin{table*}[tbhp]
\tiny
\centering
\begin{tabular}{cc|c|cccccccccc|c}
\toprule
\multicolumn{13}{c}{1D $\tb$ cross section for $\zcut = 0.2$}\\
\midrule
\multicolumn{2}{c|}{$\tb$ range} & Results & \multicolumn{10}{c|}{Uncertainty} & QED Factor\\
\cmidrule(lr){1-2}\cmidrule(lr){3-3}\cmidrule(lr){4-13}\cmidrule{14-14}
$\mathrm{Bin~Min.}$ & $\mathrm{Bin~Max.}$ & $d\sigma/d\mathrm{\tb}$ & Total & Stat. & JES & RCES & $\theta_{\mathrm{HFS}}$ & $E_e$ & $\theta_e$ & Lumi & Reg. & Model & $c_\mathrm{QED}$\\
& & $[\text{pb}]$ & $_{[\%]}$ & $_{[\%]}$ & $_{[\%]}$ & $_{[\%]}$ & $_{[\%]}$ & $_{[\%]}$ & $_{[\%]}$ & $_{[\%]}$ &  $_{[\%]}$ & $_{[\%]}$ &\\
\midrule
0.00 & 0.05 & 1.02e+04 & 6.2 & 0.5 & 1.7 & 0.1 & -0.0 & -0.9 & 0.2 & 2.7 & 2.5 & 2.0 & 1.168\\
0.05 & 0.10 & 6.10e+03 & 9.5 & 0.9 & 0.5 & 0.1 & 0.1 & -0.9 & 0.4 & 2.7 & 7.0 & 4.1 & 1.156\\
0.10 & 0.15 & 3.34e+03 & 12.0 & 1.7 & -0.2 & 0.3 & 0.1 & -1.2 & 0.4 & 2.7 & 9.8 & 4.9 & 1.155\\
0.15 & 0.22 & 1.49e+03 & 8.5 & 3.1 & 0.7 & 0.4 & 0.1 & -1.6 & 0.4 & 2.7 & 5.5 & 3.3 & 1.154\\
0.22 & 0.30 & 1.09e+03 & 9.6 & 4.2 & 0.8 & 0.3 & 0.1 & -1.6 & 0.4 & 2.7 & 4.3 & 6.0 & 1.153\\
0.30 & 0.40 & 7.19e+02 & 9.0 & 4.3 & 2.2 & -0.5 & 0.1 & -2.0 & 0.5 & 2.7 & 3.8 & 4.3 & 1.156\\
0.40 & 0.50 & 6.36e+02 & 8.0 & 5.0 & 1.7 & -0.6 & 0.1 & -1.7 & 0.4 & 2.7 & 2.9 & 2.4 & 1.157\\
0.50 & 0.60 & 5.92e+02 & 7.8 & 5.5 & 1.0 & -0.5 & 0.0 & -1.4 & 0.4 & 2.7 & 1.9 & 1.7 & 1.156\\
0.60 & 0.70 & 4.92e+02 & 7.6 & 5.7 & 0.3 & -0.5 & 0.0 & -1.0 & 0.2 & 2.7 & 1.1 & 1.6 & 1.155\\
0.70 & 0.80 & 4.44e+02 & 6.9 & 5.0 & -0.1 & -0.2 & 0.1 & -0.6 & 0.3 & 2.7 & 0.6 & 1.2 & 1.155\\
0.80 & 0.90 & 4.24e+02 & 6.7 & 4.2 & 0.6 & -1.2 & -0.3 & -0.5 & 0.2 & 2.7 & 0.3 & 2.2 & 1.157\\
0.90 & 1.00 & 5.28e+02 & 5.2 & 1.9 & -0.8 & 0.0 & 0.2 & 0.4 & 0.1 & 2.7 & 0.2 & 1.2 & 1.157\\

\bottomrule
\end{tabular}
\caption{Same as Table~\ref{tab:GIM05_1D} for the \tb cross section at $\zcut = 0.2$.}
\label{tab:GrTau2_1D} 
\end{table*}

Numerical data on double-differential measurements as a function of the groomed invariant mass, $\zcut$ and $Q^2$ are shown in Tab.~\ref{tab:GIM2D_Q1} for $150<\Qsq<200$~\GeVsq, Tab.~\ref{tab:GIM2D_Q2} for $200<\Qsq<282$~\GeVsq, Tab.~\ref{tab:GIM2D_Q3} for $282<\Qsq<447$~\GeVsq, Tab.~\ref{tab:GIM2D_Q4} for $447<\Qsq<1122$~\GeVsq, and Tab.~\ref{tab:GIM2D_Q5} for $1122<\Qsq<20000$~\GeVsq. Note that $Q^2_{\mathrm{min}}$ and hence the definition of the GIM variable is different for each $Q^2$ interval.
\begin{table*}[tbhp]
\tiny\centering\begin{tabular}{cc|cccc|cccc|cccc}
\toprule
\multicolumn{14}{c}{Groomed invariant mass cross section for three values of \zcut~and $150<\Qsq<200$~\GeVsq} \\
\midrule\multicolumn{2}{c}{ln($M^2_{\mathrm{Gr.}}/\Qsq_{\mathrm{min.}}$) Range}  & \multicolumn{4}{|c|}{$\zcut=0.05$} & \multicolumn{4}{c|}{$\zcut=0.1$} & \multicolumn{4}{c|}{$\zcut=0.2$} \\
\cmidrule(lr){3-6}\cmidrule(lr){7-10}\cmidrule(lr){11-14}
& & $d\sigma/d\mathrm{GIM}$ & $\mathrm{Stat.}$ & $\mathrm{Sys.}$ & $c_\mathrm{QED}$ & $d\sigma/d\mathrm{GIM}$ & $\mathrm{Stat.}$ & $\mathrm{Sys.}$ & $c_\mathrm{QED}$ &  $d\sigma/d\mathrm{GIM}$ & $\mathrm{Stat.}$ & $\mathrm{Sy\
s.}$ & $c_\mathrm{QED}$ \\
$\mathrm{Bin~Min.}$ & $\mathrm{Bin~Max.}$ & [\text{pb}] & [\%] & [\%] & & [\text{pb}] & [\%] & [\%] &  & [\text{pb}] & [\%] &[\%]&\\
\midrule
-9 & -4 & $2.12\cdot 10^{0}$ & 16.4 & 65.3 & 1.140 & $2.07\cdot 10^{1}$ & 6.0 & 11.3 & 1.132 & $5.86\cdot 10^{1}$ & 3.4 & 6.1 & 1.131\\
-4 & -3 & $2.38\cdot 10^{1}$ & 4.4 & 25.1 & 1.131 & $6.83\cdot 10^{1}$ & 2.0 & 10.3 & 1.132 & $1.03\cdot 10^{2}$ & 1.9 & 6.5 & 1.133\\
-3 & -2 & $7.86\cdot 10^{1}$ & 1.8 & 12.0 & 1.132 & $1.09\cdot 10^{2}$ & 1.4 & 8.5 & 1.135 & $9.54\cdot 10^{1}$ & 2.1 & 7.3 & 1.136\\
-2 & -1 & $1.10\cdot 10^{2}$ & 1.4 & 11.5 & 1.136 & $8.95\cdot 10^{1}$ & 1.6 & 8.8 & 1.137 & $5.61\cdot 10^{1}$ & 2.7 & 8.2 & 1.135\\
-1 & 0 & $8.96\cdot 10^{1}$ & 1.3 & 7.6 & 1.132 & $6.07\cdot 10^{1}$ & 1.6 & 8.6 & 1.133 & $3.60\cdot 10^{1}$ & 2.6 & 8.8 & 1.134\\
0 & 1 & $7.43\cdot 10^{1}$ & 1.4 & 9.1 & 1.135 & $5.34\cdot 10^{1}$ & 1.9 & 6.3 & 1.134 & $3.50\cdot 10^{1}$ & 3.4 & 6.9 & 1.136\\
1 & 2 & $5.08\cdot 10^{1}$ & 2.3 & 6.4 & 1.136 & $3.49\cdot 10^{1}$ & 3.0 & 6.5 & 1.135 & $1.89\cdot 10^{1}$ & 4.6 & 7.9 & 1.137\\
2 & 3 & $2.37\cdot 10^{1}$ & 4.3 & 8.2 & 1.136 & $1.47\cdot 10^{1}$ & 5.2 & 7.0 & 1.136 & $7.86\cdot 10^{0}$ & 7.4 & 7.1 & 1.137\\
3 & 4 &$7.50\cdot 10^{0}$ & 9.4 & 6.6 & 1.142 & $3.20\cdot 10^{0}$ & 11.8 & 6.8 & 1.135 & $1.22\cdot 10^{0}$ & 20.9 & 7.7 & 1.142\\
\bottomrule
\end{tabular}
\caption{Groomed invariant mass cross section in the range $150<\Qsq<200$~\GeVsq. $Q^2_{\mathrm{min.}}$ is set to 150 GeV$^{2}$. The statistical error represents the outcome of the replica method described in Sec.~\ref{sect:Uncertainties}. The systematic uncertainty is the quadrature sum of all the sources listed in Sec.~\ref{sect:Uncertainties}. The total uncertainty on the data is the quadrature sum of the statistical and systematic errors given here. $c_{\mathrm{QED}}$ is the QED correction factor as derived from HERACLES; the radiative cross section can be recovered by dividing the given data by the corresponding value of $c_{\mathrm{QED}}$. The lowest bins in groomed invariant mass were consolidated due to low statistics. The binning follows the reduced low GIM binning given in Table~\ref{tab:binnings}.                   
}
\label{tab:GIM2D_Q1}
\end{table*}

\begin{table*}[tbhp]
\tiny\centering\begin{tabular}{cc|cccc|cccc|cccc}
\toprule
\multicolumn{14}{c}{Groomed invariant mass cross section for three values of \zcut~and $200<\Qsq<282$~\GeVsq} \\
\midrule\multicolumn{2}{c}{ln($M^2_{\mathrm{Gr.}}/\Qsq_{\mathrm{min.}}$) Range}  & \multicolumn{4}{|c|}{$\zcut=0.05$} & \multicolumn{4}{c|}{$\zcut=0.1$} & \multicolumn{4}{c|}{$\zcut=0.2$} \\
\cmidrule(lr){3-6}\cmidrule(lr){7-10}\cmidrule(lr){11-14}
& & $d\sigma/d\mathrm{GIM}$ & $\mathrm{Stat.}$ & $\mathrm{Sys.}$ & $c_\mathrm{QED}$ & $d\sigma/d\mathrm{GIM}$ & $\mathrm{Stat.}$ & $\mathrm{Sys.}$ & $c_\mathrm{QED}$ &  $d\sigma/d\mathrm{GIM}$ & $\mathrm{Stat.}$ & $\mathrm{Sy\
s.}$ & $c_\mathrm{QED}$ \\
$\mathrm{Bin~Min.}$ & $\mathrm{Bin~Max.}$ & [\text{pb}] & [\%] & [\%] & & [\text{pb}] & [\%] & [\%] &  & [\text{pb}] & [\%] &[\%]&\\
\midrule
-9 & -5 & $3.46\cdot 10^{-1}$ & 29.6 & 9.6 & 1.135 & $1.20\cdot 10^{0}$ & 9.7 & 15.0 & 1.130 & $8.58\cdot 10^{0}$ & 4.6 & 9.5 & 1.132\\
-5 & -4 & $4.32\cdot 10^{0}$ & 12.3 & 49.4 & 1.125 & $2.34\cdot 10^{1}$ & 4.7 & 11.4 & 1.133 & $5.70\cdot 10^{1}$ & 3.2 & 6.4 & 1.131\\
-4 & -3 & $2.97\cdot 10^{1}$ & 3.7 & 23.0 & 1.133 & $7.14\cdot 10^{1}$ & 1.9 & 7.6 & 1.131 & $8.99\cdot 10^{1}$ & 2.1 & 6.7 & 1.134\\
-3 & -2 & $7.59\cdot 10^{1}$ & 1.7 & 9.9 & 1.132 & $8.84\cdot 10^{1}$ & 1.7 & 7.6 & 1.132 & $7.10\cdot 10^{1}$ & 2.7 & 7.0 & 1.134\\
-2 & -1 & $8.50\cdot 10^{1}$ & 1.6 & 10.2 & 1.133 & $6.38\cdot 10^{1}$ & 1.9 & 8.8 & 1.137 & $3.54\cdot 10^{1}$ & 3.0 & 10.6 & 1.134\\
-1 & 0 & $6.46\cdot 10^{1}$ & 1.4 & 10.2 & 1.135 & $4.33\cdot 10^{1}$ & 1.7 & 9.9 & 1.135 & $2.68\cdot 10^{1}$ & 3.0 & 9.1 & 1.135\\
0 & 1 & $5.92\cdot 10^{1}$ & 1.7 & 5.8 & 1.134 & $4.40\cdot 10^{1}$ & 2.2 & 7.3 & 1.134 & $2.78\cdot 10^{1}$ & 3.8 & 7.3 & 1.135\\
1 & 2 & $4.11\cdot 10^{1}$ & 2.8 & 7.9 & 1.136 & $2.86\cdot 10^{1}$ & 3.6 & 7.3 & 1.137 & $1.77\cdot 10^{1}$ & 5.3 & 7.4 & 1.139\\
2 & 3 & $1.65\cdot 10^{1}$ & 5.0 & 7.2 & 1.137 & $9.11\cdot 10^{0}$ & 6.2 & 6.8 & 1.140 & $4.55\cdot 10^{0}$ & 8.2 & 7.4 & 1.139\\
3 & 4 & $4.59\cdot 10^{0}$ & 11.0 & 6.8 & 1.140 & $2.09\cdot 10^{0}$ & 14.3 & 10.6 & 1.136 & $9.37\cdot 10^{-1}$ & 25.4 & 14.9 & 1.137\\
\bottomrule
\end{tabular}
\caption{                                                                                                                                                                                                                                                                                      
Same as Table \ref{tab:GIM2D_Q1} for groomed invariant mass cross sections in the range $200<\Qsq<282$~\GeVsq. $Q^2_{\mathrm{min.}}$ is set to 200 GeV$^{2}$. The binning follows the standard GIM binning given in Table~\ref{tab:binnings}.                                                                                                                                                                                                                   
}
\label{tab:GIM2D_Q2}
\end{table*}

\begin{table*}[tbhp]
\tiny\centering\begin{tabular}{cc|cccc|cccc|cccc}
\toprule
\multicolumn{14}{c}{Groomed invariant mass cross section for three values of \zcut~and $282<\Qsq<447$~\GeVsq} \\
\midrule\multicolumn{2}{c}{ln($M^2_{\mathrm{Gr.}}/\Qsq_{\mathrm{min.}}$) Range}  & \multicolumn{4}{|c|}{$\zcut=0.05$} & \multicolumn{4}{c|}{$\zcut=0.1$} & \multicolumn{4}{c|}{$\zcut=0.2$} \\
\cmidrule(lr){3-6}\cmidrule(lr){7-10}\cmidrule(lr){11-14}
& & $d\sigma/d\mathrm{GIM}$ & $\mathrm{Stat.}$ & $\mathrm{Sys.}$ & $c_\mathrm{QED}$ & $d\sigma/d\mathrm{GIM}$ & $\mathrm{Stat.}$ & $\mathrm{Sys.}$ & $c_\mathrm{QED}$ &  $d\sigma/d\mathrm{GIM}$ & $\mathrm{Stat.}$ & $\mathrm{Sy\
s.}$ & $c_\mathrm{QED}$ \\
$\mathrm{Bin~Min.}$ & $\mathrm{Bin~Max.}$ & [\text{pb}] & [\%] & [\%] & & [\text{pb}] & [\%] & [\%] &  & [\text{pb}] & [\%] &[\%]&\\
\midrule
-9 & -5 & $2.75\cdot 10^{-1}$ & 18.3 & 26.3 & 1.134 & $2.39\cdot 10^{0}$ & 7.4 & 10.5 & 1.134 & $1.12\cdot 10^{1}$ & 3.6 & 6.7 & 1.136\\
-5 & -4 & $7.86\cdot 10^{0}$ & 8.8 & 47.9 & 1.137 & $3.09\cdot 10^{1}$ & 3.8 & 11.1 & 1.136 & $5.66\cdot 10^{1}$ & 3.0 & 6.0 & 1.137\\
-4 & -3 & $3.52\cdot 10^{1}$ & 3.1 & 23.8 & 1.137 & $6.59\cdot 10^{1}$ & 2.1 & 7.5 & 1.135 & $7.05\cdot 10^{1}$ & 2.5 & 6.3 & 1.134\\
 -3 & -2 & $7.06\cdot 10^{1}$ & 2.0 & 10.3 & 1.134 & $6.86\cdot 10^{1}$ & 2.0 & 8.2 & 1.135 & $4.71\cdot 10^{1}$ & 3.3 & 8.3 & 1.135\\
-2 & -1 & $6.64\cdot 10^{1}$ & 1.7 & 10.1 & 1.135 & $4.49\cdot 10^{1}$ & 2.1 & 9.6 & 1.136 & $2.76\cdot 10^{1}$ & 3.3 & 10.3 & 1.137\\
-1 & 0 & $4.89\cdot 10^{1}$ & 1.7 & 9.4 & 1.137 & $3.39\cdot 10^{1}$ & 2.2 & 8.0 & 1.137 & $1.98\cdot 10^{1}$ & 3.4 & 8.3 & 1.138\\
0 & 1 & $4.28\cdot 10^{1}$ & 2.2 & 6.4 & 1.138 & $3.14\cdot 10^{1}$ & 2.9 & 7.2 & 1.137 & $2.09\cdot 10^{1}$ & 4.6 & 7.3 & 1.137\\
1 & 2 & $2.85\cdot 10^{1}$ & 3.5 & 7.5 & 1.136 & $2.01\cdot 10^{1}$ & 4.3 & 7.0 & 1.138 & $1.29\cdot 10^{1}$ & 6.5 & 7.3 & 1.140\\
2 & 3 & $1.19\cdot 10^{1}$ & 6.1 & 6.1 & 1.141 & $7.11\cdot 10^{0}$ & 7.9 & 5.4 & 1.141 & $3.45\cdot 10^{0}$ & 10.8 & 5.9 & 1.135\\
3 & 4 & $2.64\cdot 10^{0}$ & 15.7 & 8.9 & 1.157 & $1.02\cdot 10^{0}$ & 23.0 & 9.9 & 1.150 & $3.07\cdot 10^{-1}$ & 36.3 & 14.8 & 1.148\\
\bottomrule
\end{tabular}
\caption{
Same as Table \ref{tab:GIM2D_Q1} for groomed invariant mass cross sections in the range $282<\Qsq<447$~\GeVsq. $Q^2_{\mathrm{min.}}$ is set to 282 GeV$^{2}$. The binning follows the standard GIM binning given in Table~\ref{tab:binnings}. 
}
\label{tab:GIM2D_Q3}
\end{table*}

\begin{table*}[tbhp]
\tiny\centering\begin{tabular}{cc|cccc|cccc|cccc}
\toprule
\multicolumn{14}{c}{Groomed invariant mass cross section for three values of \zcut~and $447<\Qsq<1122$~\GeVsq} \\
\midrule\multicolumn{2}{c}{ln($M^2_{\mathrm{Gr.}}/\Qsq_{\mathrm{min.}}$) Range}  & \multicolumn{4}{|c|}{$\zcut=0.05$} & \multicolumn{4}{c|}{$\zcut=0.1$} & \multicolumn{4}{c|}{$\zcut=0.2$} \\
\cmidrule(lr){3-6}\cmidrule(lr){7-10}\cmidrule(lr){11-14}
& & $d\sigma/d\mathrm{GIM}$ & $\mathrm{Stat.}$ & $\mathrm{Sys.}$ & $c_\mathrm{QED}$ & $d\sigma/d\mathrm{GIM}$ & $\mathrm{Stat.}$ & $\mathrm{Sys.}$ & $c_\mathrm{QED}$ &  $d\sigma/d\mathrm{GIM}$ & $\mathrm{Stat.}$ & $\mathrm{Sy\
s.}$ & $c_\mathrm{QED}$ \\
$\mathrm{Bin~Min.}$ & $\mathrm{Bin~Max.}$ & [\text{pb}] & [\%] & [\%] & & [\text{pb}] & [\%] & [\%] &  & [\text{pb}] & [\%] &[\%]&\\
\midrule
-9 & -5 & $5.48\cdot 10^{-1}$ & 11.6 & 11.6 & 1.141 & $3.54\cdot 10^{0}$ & 4.8 & 13.7 & 1.145 & $1.33\cdot 10^{1}$ & 2.7 & 9.3 & 1.142\\
-5 & -4 & $1.41\cdot 10^{1}$ & 5.5 & 46.4 & 1.144 & $4.29\cdot 10^{1}$ & 2.9 & 10.7 & 1.142 & $6.21\cdot 10^{1}$ & 3.1 & 5.0 & 1.141\\
-4 & -3 & $4.54\cdot 10^{1}$ & 2.7 & 21.9 & 1.140 & $6.31\cdot 10^{1}$ & 2.2 & 5.9 & 1.141 & $5.71\cdot 10^{1}$ & 3.3 & 5.3 & 1.142\\
-3 & -2 & $6.21\cdot 10^{1}$ & 2.1 & 9.7 & 1.143 & $5.05\cdot 10^{1}$ & 2.5 & 7.0 & 1.144 & $3.28\cdot 10^{1}$ & 3.8 & 9.1 & 1.140\\
-2 & -1 & $5.21\cdot 10^{1}$ & 1.9 & 9.8 & 1.142 & $3.44\cdot 10^{1}$ & 2.3 & 9.4 & 1.142 & $1.93\cdot 10^{1}$ & 3.5 & 9.0 & 1.143\\
-1 & 0 & $4.14\cdot 10^{1}$ & 1.9 & 6.9 & 1.145 & $3.00\cdot 10^{1}$ & 2.5 & 6.2 & 1.145 & $1.95\cdot 10^{1}$ & 3.6 & 6.4 & 1.147\\
0 & 1 & $3.48\cdot 10^{1}$ & 2.5 & 5.2 & 1.144 & $2.52\cdot 10^{1}$ & 3.0 & 5.8 & 1.143 & $1.58\cdot 10^{1}$ & 4.5 & 6.0 & 1.144\\
1 & 2 & $2.13\cdot 10^{1}$ & 3.6 & 5.4 & 1.143 & $1.42\cdot 10^{1}$ & 4.4 & 4.7 & 1.144 & $8.26\cdot 10^{0}$ & 6.3 & 5.1 & 1.147\\
2 & 4 & $7.02\cdot 10^{0}$ & 6.5 & 3.6 & 1.150 & $4.52\cdot 10^{0}$ & 8.3 & 4.6 & 1.149 & $2.21\cdot 10^{0}$ & 11.8 & 6.2 & 1.147\\
\bottomrule
\end{tabular}
\caption{
Same as Table \ref{tab:GIM2D_Q1} for groomed invariant mass cross sections in the range $447<\Qsq<1122$~\GeVsq. $Q^2_{\mathrm{min.}}$ is set to 447 GeV$^{2}$. The binning follows the reduced high GIM binning given in Table~\ref{tab:binnings}. 
}
\label{tab:GIM2D_Q4}
\end{table*}

\begin{table*}[tbhp]
\tiny\centering\begin{tabular}{cc|cccc|cccc|cccc}
\toprule
\multicolumn{14}{c}{Groomed invariant mass cross section for three values of \zcut~and $1122<\Qsq<20000$~\GeVsq} \\
\midrule\multicolumn{2}{c}{ln($M^2_{\mathrm{Gr.}}/\Qsq_{\mathrm{min.}}$) Range}  & \multicolumn{4}{|c|}{$\zcut=0.05$} & \multicolumn{4}{c|}{$\zcut=0.1$} & \multicolumn{4}{c|}{$\zcut=0.2$} \\
\cmidrule(lr){3-6}\cmidrule(lr){7-10}\cmidrule(lr){11-14}
& & $d\sigma/d\mathrm{GIM}$ & $\mathrm{Stat.}$ & $\mathrm{Sys.}$ & $c_\mathrm{QED}$ & $d\sigma/d\mathrm{GIM}$ & $\mathrm{Stat.}$ & $\mathrm{Sys.}$ & $c_\mathrm{QED}$ &  $d\sigma/d\mathrm{GIM}$ & $\mathrm{Stat.}$ & $\mathrm{Sy\
s.}$ & $c_\mathrm{QED}$ \\
$\mathrm{Bin~Min.}$ & $\mathrm{Bin~Max.}$ & [\text{pb}] & [\%] & [\%] & & [\text{pb}] & [\%] & [\%] &  & [\text{pb}] & [\%] &[\%]&\\
\midrule
-9 & -5 & $1.01\cdot 10^{0}$ & 8.8 & 15.8 & 1.210 & $4.70\cdot 10^{0}$ & 4.6 & 16.7 & 1.207 & $1.03\cdot 10^{1}$ & 4.0 & 12.3 & 1.197\\
-5 & -4 & $1.58\cdot 10^{1}$ & 6.1 & 45.3 & 1.210 & $2.50\cdot 10^{1}$ & 5.2 & 12.8 & 1.210 & $2.44\cdot 10^{1}$ & 6.7 & 7.4 & 1.209\\
-4 & -3 & $2.45\cdot 10^{1}$ & 4.8 & 20.8 & 1.210 & $2.13\cdot 10^{1}$ & 4.7 & 10.1 & 1.210 & $1.64\cdot 10^{1}$ & 6.3 & 12.1 & 1.211\\
-3 & -2 & $2.21\cdot 10^{1}$ & 3.6 & 14.6 & 1.216 & $1.54\cdot 10^{1}$ & 3.9 & 10.5 & 1.214 & $8.47\cdot 10^{0}$ & 5.6 & 8.4 & 1.210\\
-2 & -1 & $1.69\cdot 10^{1}$ & 3.7 & 8.9 & 1.205 & $1.18\cdot 10^{1}$ & 4.2 & 7.0 & 1.204 & $7.33\cdot 10^{0}$ & 5.2 & 6.9 & 1.204\\
-1 & 0 & $1.47\cdot 10^{1}$ & 3.7 & 6.3 & 1.199 & $1.10\cdot 10^{1}$ & 4.4 & 6.5 & 1.200 & $7.21\cdot 10^{0}$ & 5.5 & 6.6 & 1.201\\
0 & 1 & $1.13\cdot 10^{1}$ & 4.4 & 5.1 & 1.204 & $7.91\cdot 10^{0}$ & 5.0 & 5.9 & 1.205 & $4.71\cdot 10^{0}$ & 6.9 & 5.6 & 1.205\\
1 & 2 & $5.20\cdot 10^{0}$ & 7.3 & 4.1 & 1.221 & $3.68\cdot 10^{0}$ & 9.1 & 4.2 & 1.224 & $2.11\cdot 10^{0}$ & 13.5 & 5.2 & 1.231\\
2 & 4 & $7.65\cdot 10^{-1}$ & 34.8 & 8.4 & 1.250 & $2.68\cdot 10^{-1}$ & 42.7 & 92.3 & 1.246 & $9.70\cdot 10^{-2}$ & 47.7 & 52.2 & 1.247\\
\bottomrule
\end{tabular}
\caption{
Same as Table \ref{tab:GIM2D_Q1} for groomed invariant mass cross sections in the range $1122<\Qsq<20000$~\GeVsq. $Q^2_{\mathrm{min.}}$ is set to 1122 GeV$^{2}$. The binning follows the reduced high GIM binning given in Table~\ref{tab:binnings}. 
}
\label{tab:GIM2D_Q5}
\end{table*}

\begin{table*}[tbhp]
\tiny\centering\begin{tabular}{cc|cccc|cccc|cccc}
\toprule
\multicolumn{14}{c}{Groomed $\tb$ cross section for three values of \zcut~and $150<\Qsq<200$~\GeVsq} \\
\midrule\multicolumn{2}{c}{\tb Range}  & \multicolumn{4}{|c|}{$\zcut=0.05$} & \multicolumn{4}{c|}{$\zcut=0.1$} & \multicolumn{4}{c|}{$\zcut=0.2$} \\
\cmidrule(lr){3-6}\cmidrule(lr){7-10}\cmidrule(lr){11-14}
& & $d\sigma/d\tb$ & $\mathrm{Stat.}$ & $\mathrm{Sys.}$ & $c_\mathrm{QED}$ & $d\sigma/d\tb$ & $\mathrm{Stat.}$ & $\mathrm{Sys.}$ & $c_\mathrm{QED}$ &  $d\sigma/d\mathrm{GIM}$ & $\mathrm{Stat.}$ & $\mathrm{Sy\
s.}$ & $c_\mathrm{QED}$ \\
$\mathrm{Bin~Min.}$ & $\mathrm{Bin~Max.}$ & [\text{pb}] & [\%] & [\%] & & [\text{pb}] & [\%] & [\%] &  & [\text{pb}] & [\%] &[\%]&\\
\midrule
0.00 & 0.05 & $4.04\cdot 10^{2}$ & 5.6 & 8.0 & 1.137 & $1.17\cdot 10^{3}$ & 3.0 & 7.1 & 1.133 & $2.26\cdot 10^{3}$ & 2.8 & 12.3 & 1.134\\
0.05 & 0.10 & $9.47\cdot 10^{2}$ & 2.9 & 8.0 & 1.128 & $1.52\cdot 10^{3}$ & 2.5 & 7.0 & 1.133 & $2.11\cdot 10^{3}$ & 2.9 & 10.6 & 1.134\\
0.10 & 0.15 & $1.26\cdot 10^{3}$ & 2.2 & 6.5 & 1.134 & $1.37\cdot 10^{3}$ & 3.1 & 7.7 & 1.132 & $1.32\cdot 10^{3}$ & 4.0 & 9.2 & 1.134\\
0.15 & 0.22 & $1.09\cdot 10^{3}$ & 2.2 & 6.8 & 1.136 & $9.00\cdot 10^{2}$ & 3.6 & 7.0 & 1.137 & $6.53\cdot 10^{2}$ & 4.9 & 8.5 & 1.134\\
0.22 & 0.30 & $7.38\cdot 10^{2}$ & 2.3 & 6.3 & 1.134 & $5.67\cdot 10^{2}$ & 3.8 & 6.8 & 1.136 & $3.81\cdot 10^{2}$ & 5.7 & 9.3 & 1.134\\
0.30 & 0.40 & $4.82\cdot 10^{2}$ & 2.0 & 7.2 & 1.134 & $3.61\cdot 10^{2}$ & 3.1 & 8.0 & 1.135 & $3.12\cdot 10^{2}$ & 5.5 & 15.9 & 1.135\\
0.40 & 0.50 & $3.63\cdot 10^{2}$ & 2.4 & 7.0 & 1.135 & $2.94\cdot 10^{2}$ & 3.9 & 7.5 & 1.134 & $2.32\cdot 10^{2}$ & 6.5 & 13.5 & 1.137\\
0.50 & 0.60 & $2.83\cdot 10^{2}$ & 3.3 & 6.3 & 1.134 & $2.64\cdot 10^{2}$ & 4.8 & 5.9 & 1.133 & $2.49\cdot 10^{2}$ & 7.7 & 13.2 & 1.135\\
0.60 & 0.70 & $2.40\cdot 10^{2}$ & 4.1 & 5.6 & 1.134 & $2.30\cdot 10^{2}$ & 5.8 & 5.3 & 1.134 & $1.95\cdot 10^{2}$ & 8.6 & 7.8 & 1.133\\
0.70 & 0.80 & $1.89\cdot 10^{2}$ & 4.8 & 5.8 & 1.133 & $1.76\cdot 10^{2}$ & 6.5 & 6.9 & 1.132 & $1.65\cdot 10^{2}$ & 8.4 & 7.7 & 1.133\\
0.80 & 0.90 & $1.98\cdot 10^{2}$ & 5.3 & 6.3 & 1.139 & $2.05\cdot 10^{2}$ & 5.7 & 6.8 & 1.139 & $1.57\cdot 10^{2}$ & 7.5 & 6.9 & 1.141\\
0.90 & 1.00 & $3.17\cdot 10^{2}$ & 2.7 & 5.8 & 1.136 & $2.66\cdot 10^{2}$ & 3.2 & 6.1 & 1.136 & $1.93\cdot 10^{2}$ & 3.7 & 5.5 & 1.136\\
\bottomrule
\end{tabular}
        \caption{
Same as Table \ref{tab:GIM2D_Q1} for \tb cross sections in the range $150<\Qsq<200$~\GeVsq. The binning follows the standard \tb binning given in Table~\ref{tab:binnings}. 
}      
\label{tab:GrTau2D_Q1}
\end{table*}

\begin{table*}[tbhp]
\tiny\centering\begin{tabular}{cc|cccc|cccc|cccc}
\toprule
\multicolumn{14}{c}{Groomed $\tb$ cross section for three values of \zcut~and $200<\Qsq<282$~\GeVsq} \\
\midrule\multicolumn{2}{c}{\tb Range}  & \multicolumn{4}{|c|}{$\zcut=0.05$} & \multicolumn{4}{c|}{$\zcut=0.1$} & \multicolumn{4}{c|}{$\zcut=0.2$} \\
\cmidrule(lr){3-6}\cmidrule(lr){7-10}\cmidrule(lr){11-14}
& & $d\sigma/d\tb$ & $\mathrm{Stat.}$ & $\mathrm{Sys.}$ & $c_\mathrm{QED}$ & $d\sigma/d\tb$ & $\mathrm{Stat.}$ & $\mathrm{Sys.}$ & $c_\mathrm{QED}$ &  $d\sigma/d\mathrm{GIM}$ & $\mathrm{Stat.}$ & $\mathrm{Sy\
s.}$ & $c_\mathrm{QED}$ \\
$\mathrm{Bin~Min.}$ & $\mathrm{Bin~Max.}$ & [\text{pb}] & [\%] & [\%] & & [\text{pb}] & [\%] & [\%] &  & [\text{pb}] & [\%] &[\%]&\\
\midrule
0.00 & 0.05 & $5.73\cdot 10^{2}$ & 3.3 & 7.4 & 1.132 & $1.40\cdot 10^{3}$ & 2.3 & 5.4 & 1.130 & $2.46\cdot 10^{3}$ & 2.3 & 12.3 & 1.131\\
0.05 & 0.10 & $1.08\cdot 10^{3}$ & 2.2 & 6.5 & 1.132 & $1.50\cdot 10^{3}$ & 2.5 & 6.5 & 1.134 & $1.81\cdot 10^{3}$ & 2.9 & 11.1 & 1.133\\
0.10 & 0.15 & $1.15\cdot 10^{3}$ & 2.6 & 7.0 & 1.132 & $1.07\cdot 10^{3}$ & 3.9 & 6.6 & 1.132 & $9.09\cdot 10^{2}$ & 5.2 & 8.7 & 1.135\\
0.15 & 0.22 & $8.16\cdot 10^{2}$ & 2.7 & 6.2 & 1.134 & $6.39\cdot 10^{2}$ & 4.1 & 7.0 & 1.134 & $4.69\cdot 10^{2}$ & 5.7 & 9.8 & 1.133\\
0.22 & 0.30 & $5.02\cdot 10^{2}$ & 2.5 & 7.6 & 1.136 & $3.87\cdot 10^{2}$ & 4.1 & 8.7 & 1.138 & $2.93\cdot 10^{2}$ & 6.9 & 11.7 & 1.136\\
0.30 & 0.40 & $3.58\cdot 10^{2}$ & 2.4 & 6.7 & 1.135 & $2.74\cdot 10^{2}$ & 3.7 & 7.2 & 1.134 & $2.06\cdot 10^{2}$ & 7.2 & 9.7 & 1.136\\
0.40 & 0.50 & $2.73\cdot 10^{2}$ & 3.2 & 6.7 & 1.131 & $2.33\cdot 10^{2}$ & 4.9 & 7.4 & 1.131 & $1.96\cdot 10^{2}$ & 7.9 & 13.6 & 1.134\\
0.50 & 0.60 & $2.12\cdot 10^{2}$ & 4.1 & 6.8 & 1.136 & $1.90\cdot 10^{2}$ & 6.0 & 6.7 & 1.137 & $1.60\cdot 10^{2}$ & 9.2 & 8.8 & 1.134\\
0.60 & 0.70 & $1.83\cdot 10^{2}$ & 5.6 & 6.4 & 1.137 & $1.66\cdot 10^{2}$ & 7.4 & 6.3 & 1.138 & $1.47\cdot 10^{2}$ & 10.4 & 11.4 & 1.138\\
0.70 & 0.80 & $1.44\cdot 10^{2}$ & 6.3 & 6.7 & 1.134 & $1.37\cdot 10^{2}$ & 7.8 & 6.2 & 1.135 & $1.20\cdot 10^{2}$ & 10.3 & 6.2 & 1.136\\
0.80 & 0.90 & $1.56\cdot 10^{2}$ & 6.8 & 7.0 & 1.136 & $1.62\cdot 10^{2}$ & 7.0 & 6.8 & 1.136 & $1.29\cdot 10^{2}$ & 8.6 & 7.2 & 1.135\\
0.90 & 1.00 & $2.22\cdot 10^{2}$ & 3.3 & 5.3 & 1.139 & $1.86\cdot 10^{2}$ & 3.7 & 5.5 & 1.139 & $1.49\cdot 10^{2}$ & 4.4 & 5.7 & 1.138\\
\bottomrule
\end{tabular}
\caption{                                                                                                             Same as Table \ref{tab:GIM2D_Q1} for \tb cross sections in the range $200<\Qsq<282$~\GeVsq. The binning follows the standard \tb binning given in Table~\ref{tab:binnings}.                                                                                                             
}
\label{tab:GrTau2D_Q2}
\end{table*}

\begin{table*}[tbhp]
\tiny\centering\begin{tabular}{cc|cccc|cccc|cccc}
\toprule
\multicolumn{14}{c}{Groomed $\tb$ cross section for three values of \zcut~and $282<\Qsq<447$~\GeVsq} \\
\midrule\multicolumn{2}{c}{\tb Range}  & \multicolumn{4}{|c|}{$\zcut=0.05$} & \multicolumn{4}{c|}{$\zcut=0.1$} & \multicolumn{4}{c|}{$\zcut=0.2$} \\
\cmidrule(lr){3-6}\cmidrule(lr){7-10}\cmidrule(lr){11-14}
& & $d\sigma/d\tb$ & $\mathrm{Stat.}$ & $\mathrm{Sys.}$ & $c_\mathrm{QED}$ & $d\sigma/d\tb$ & $\mathrm{Stat.}$ & $\mathrm{Sys.}$ & $c_\mathrm{QED}$ &  $d\sigma/d\mathrm{GIM}$ & $\mathrm{Stat.}$ & $\mathrm{Sy\
s.}$ & $c_\mathrm{QED}$ \\
$\mathrm{Bin~Min.}$ & $\mathrm{Bin~Max.}$ & [\text{pb}] & [\%] & [\%] & & [\text{pb}] & [\%] & [\%] &  & [\text{pb}] & [\%] &[\%]&\\
\midrule
0.00 & 0.05 & $7.86\cdot 10^{2}$ & 2.3 & 8.5 & 1.136 & $1.70\cdot 10^{3}$ & 1.8 & 5.4 & 1.134 & $2.68\cdot 10^{3}$ & 2.2 & 11.4 & 1.135\\
0.05 & 0.10 & $1.17\cdot 10^{3}$ & 2.3 & 6.0 & 1.134 & $1.34\cdot 10^{3}$ & 2.9 & 5.8 & 1.135 & $1.45\cdot 10^{3}$ & 3.5 & 10.9 & 1.137\\
0.10 & 0.15 & $9.89\cdot 10^{2}$ & 3.4 & 7.0 & 1.133 & $7.91\cdot 10^{2}$ & 4.9 & 6.6 & 1.137 & $6.00\cdot 10^{2}$ & 6.0 & 9.0 & 1.137\\
0.15 & 0.22 & $5.68\cdot 10^{2}$ & 2.9 & 8.6 & 1.135 & $4.29\cdot 10^{2}$ & 4.7 & 8.3 & 1.139 & $3.34\cdot 10^{2}$ & 7.0 & 10.0 & 1.137\\
0.22 & 0.30 & $3.84\cdot 10^{2}$ & 2.7 & 7.0 & 1.137 & $2.86\cdot 10^{2}$ & 4.8 & 6.7 & 1.137 & $2.43\cdot 10^{2}$ & 7.9 & 9.7 & 1.135\\
0.30 & 0.40 & $2.59\cdot 10^{2}$ & 2.9 & 6.0 & 1.139 & $2.03\cdot 10^{2}$ & 4.7 & 6.4 & 1.138 & $1.44\cdot 10^{2}$ & 8.2 & 9.4 & 1.134\\
0.40 & 0.50 & $1.89\cdot 10^{2}$ & 4.3 & 6.5 & 1.138 & $1.60\cdot 10^{2}$ & 5.7 & 6.5 & 1.137 & $1.17\cdot 10^{2}$ & 9.9 & 11.2 & 1.143\\
0.50 & 0.60 & $1.48\cdot 10^{2}$ & 5.8 & 6.3 & 1.138 & $1.36\cdot 10^{2}$ & 7.4 & 6.6 & 1.138 & $1.10\cdot 10^{2}$ & 11.0 & 10.7 & 1.139\\
0.60 & 0.70 & $1.28\cdot 10^{2}$ & 7.4 & 6.2 & 1.139 & $1.16\cdot 10^{2}$ & 9.3 & 5.6 & 1.139 & $9.86\cdot 10^{1}$ & 11.6 & 8.7 & 1.138\\
0.70 & 0.80 & $1.02\cdot 10^{2}$ & 8.6 & 5.5 & 1.139 & $1.00\cdot 10^{2}$ & 9.6 & 5.7 & 1.137 & $1.08\cdot 10^{2}$ & 11.9 & 7.2 & 1.137\\
0.80 & 0.90 & $9.47\cdot 10^{1}$ & 8.7 & 6.6 & 1.135 & $1.02\cdot 10^{2}$ & 8.9 & 6.2 & 1.136 & $8.05\cdot 10^{1}$ & 10.6 & 6.4 & 1.135\\
0.90 & 1.00 & $1.59\cdot 10^{2}$ & 3.9 & 5.2 & 1.138 & $1.37\cdot 10^{2}$ & 4.3 & 5.3 & 1.138 & $1.10\cdot 10^{2}$ & 5.1 & 5.0 & 1.143\\
\bottomrule
\end{tabular}
\caption{                                                                                                             Same as Table \ref{tab:GIM2D_Q1} for \tb cross sections in the range $282<\Qsq<447$~\GeVsq. The binning follows the standard \tb binning given in Table~\ref{tab:binnings}.                                                   
}
\label{tab:GrTau2D_Q3}
\end{table*}

\begin{table*}[tbhp]
\tiny\centering\begin{tabular}{cc|cccc|cccc|cccc}
\toprule
\multicolumn{14}{c}{Groomed $\tb$ cross section for three values of \zcut~and $447<\Qsq<1122$~\GeVsq} \\
\midrule\multicolumn{2}{c}{\tb Range}  & \multicolumn{4}{|c|}{$\zcut=0.05$} & \multicolumn{4}{c|}{$\zcut=0.1$} & \multicolumn{4}{c|}{$\zcut=0.2$} \\
\cmidrule(lr){3-6}\cmidrule(lr){7-10}\cmidrule(lr){11-14}
& & $d\sigma/d\tb$ & $\mathrm{Stat.}$ & $\mathrm{Sys.}$ & $c_\mathrm{QED}$ & $d\sigma/d\tb$ & $\mathrm{Stat.}$ & $\mathrm{Sys.}$ & $c_\mathrm{QED}$ &  $d\sigma/d\mathrm{GIM}$ & $\mathrm{Stat.}$ & $\mathrm{Sy\
s.}$ & $c_\mathrm{QED}$ \\
$\mathrm{Bin~Min.}$ & $\mathrm{Bin~Max.}$ & [\text{pb}] & [\%] & [\%] & & [\text{pb}] & [\%] & [\%] &  & [\text{pb}] & [\%] &[\%]&\\
\midrule
0.00 & 0.05 & $1.35\cdot 10^{3}$ & 1.6 & 8.4 & 1.143 & $2.26\cdot 10^{3}$ & 1.5 & 5.4 & 1.143 & $3.15\cdot 10^{3}$ & 2.1 & 11.0 & 1.142\\
0.05 & 0.10 & $1.20\cdot 10^{3}$ & 2.6 & 10.6 & 1.142 & $1.16\cdot 10^{3}$ & 3.3 & 7.5 & 1.141 & $1.16\cdot 10^{3}$ & 3.6 & 12.1 & 1.140\\
0.10 & 0.15 & $7.32\cdot 10^{2}$ & 3.3 & 10.5 & 1.144 & $5.15\cdot 10^{2}$ & 5.1 & 11.1 & 1.144 & $3.78\cdot 10^{2}$ & 7.7 & 6.9 & 1.145\\
0.15 & 0.22 & $3.95\cdot 10^{2}$ & 3.1 & 6.3 & 1.140 & $2.90\cdot 10^{2}$ & 5.3 & 7.2 & 1.141 & $2.24\cdot 10^{2}$ & 7.5 & 10.0 & 1.143\\
0.22 & 0.30 & $2.99\cdot 10^{2}$ & 3.5 & 6.1 & 1.143 & $2.34\cdot 10^{2}$ & 5.9 & 6.1 & 1.141 & $1.64\cdot 10^{2}$ & 8.9 & 8.7 & 1.141\\
0.30 & 0.40 & $2.12\cdot 10^{2}$ & 4.0 & 5.6 & 1.143 & $1.64\cdot 10^{2}$ & 6.0 & 6.0 & 1.144 & $1.12\cdot 10^{2}$ & 9.0 & 8.2 & 1.143\\
0.40 & 0.50 & $1.34\cdot 10^{2}$ & 5.9 & 6.0 & 1.145 & $1.26\cdot 10^{2}$ & 7.1 & 5.9 & 1.145 & $1.06\cdot 10^{2}$ & 10.7 & 11.1 & 1.148\\
0.50 & 0.60 & $1.12\cdot 10^{2}$ & 7.5 & 5.9 & 1.148 & $9.96\cdot 10^{1}$ & 9.3 & 5.4 & 1.147 & $7.10\cdot 10^{1}$ & 12.6 & 8.7 & 1.148\\
0.60 & 0.70 & $7.74\cdot 10^{1}$ & 9.7 & 5.7 & 1.144 & $7.04\cdot 10^{1}$ & 11.0 & 4.9 & 1.145 & $7.13\cdot 10^{1}$ & 13.7 & 9.6 & 1.145\\
0.70 & 0.80 & $5.83\cdot 10^{1}$ & 10.6 & 5.0 & 1.144 & $5.64\cdot 10^{1}$ & 11.5 & 5.3 & 1.144 & $4.26\cdot 10^{1}$ & 13.8 & 6.2 & 1.143\\
0.80 & 0.90 & $5.82\cdot 10^{1}$ & 11.5 & 6.3 & 1.146 & $5.65\cdot 10^{1}$ & 11.9 & 6.1 & 1.146 & $5.04\cdot 10^{1}$ & 12.6 & 6.8 & 1.149\\
0.90 & 1.00 & $9.20\cdot 10^{1}$ & 4.9 & 4.8 & 1.149 & $8.53\cdot 10^{1}$ & 5.2 & 5.1 & 1.155 & $7.24\cdot 10^{1}$ & 5.5 & 5.5 & 1.176\\
\bottomrule
\end{tabular}
\caption{                                                                                                             Same as Table \ref{tab:GIM2D_Q1} for \tb cross sections in the range $447<\Qsq<1122$~\GeVsq. The binning follows the standard \tb binning given in Table~\ref{tab:binnings}.                                                 
}
\label{tab:GrTau2D_Q4}
\end{table*}

\begin{table*}[tbhp]
\tiny\centering\begin{tabular}{cc|cccc|cccc|cccc}
\toprule
\multicolumn{14}{c}{Groomed $\tb$ cross section for three values of \zcut~and $1122<\Qsq<20000$~\GeVsq} \\
\midrule\multicolumn{2}{c}{\tb Range}  & \multicolumn{4}{|c|}{$\zcut=0.05$} & \multicolumn{4}{c|}{$\zcut=0.1$} & \multicolumn{4}{c|}{$\zcut=0.2$} \\
\cmidrule(lr){3-6}\cmidrule(lr){7-10}\cmidrule(lr){11-14}
& & $d\sigma/d\tb$ & $\mathrm{Stat.}$ & $\mathrm{Sys.}$ & $c_\mathrm{QED}$ & $d\sigma/d\tb$ & $\mathrm{Stat.}$ & $\mathrm{Sys.}$ & $c_\mathrm{QED}$ &  $d\sigma/d\mathrm{GIM}$ & $\mathrm{Stat.}$ & $\mathrm{Sy\
s.}$ & $c_\mathrm{QED}$ \\
$\mathrm{Bin~Min.}$ & $\mathrm{Bin~Max.}$ & [\text{pb}] & [\%] & [\%] & & [\text{pb}] & [\%] & [\%] &  & [\text{pb}] & [\%] &[\%]&\\
\midrule
0.00 & 0.05 & $1.10\cdot 10^{3}$ & 1.5 & 8.1 & 1.223 & $1.48\cdot 10^{3}$ & 1.6 & 8.2 & 1.220 & $1.84\cdot 10^{3}$ & 2.4 & 12.6 & 1.217\\
0.05 & 0.10 & $4.47\cdot 10^{2}$ & 4.0 & 14.3 & 1.206 & $3.28\cdot 10^{2}$ & 4.4 & 14.0 & 1.200 & $2.94\cdot 10^{2}$ & 5.1 & 13.4 & 1.192\\
0.10 & 0.15 & $1.88\cdot 10^{2}$ & 7.4 & 12.8 & 1.199 & $1.44\cdot 10^{2}$ & 10.2 & 6.0 & 1.198 & $9.94\cdot 10^{1}$ & 14.8 & 10.6 & 1.188\\
0.15 & 0.22 & $1.25\cdot 10^{2}$ & 7.6 & 6.4 & 1.194 & $8.96\cdot 10^{1}$ & 9.2 & 7.2 & 1.199 & $6.28\cdot 10^{1}$ & 14.2 & 7.7 & 1.201\\
0.22 & 0.30 & $9.64\cdot 10^{1}$ & 9.7 & 6.4 & 1.188 & $7.93\cdot 10^{1}$ & 11.9 & 6.1 & 1.191 & $4.42\cdot 10^{1}$ & 18.9 & 9.3 & 1.190\\
0.30 & 0.40 & $5.01\cdot 10^{1}$ & 12.0 & 5.3 & 1.188 & $4.25\cdot 10^{1}$ & 13.5 & 6.0 & 1.188 & $3.90\cdot 10^{1}$ & 20.3 & 7.9 & 1.188\\
0.40 & 0.50 & $3.12\cdot 10^{1}$ & 18.2 & 5.7 & 1.177 & $3.04\cdot 10^{1}$ & 21.1 & 5.3 & 1.175 & $1.73\cdot 10^{1}$ & 25.6 & 9.2 & 1.180\\
0.50 & 0.60 & $2.47\cdot 10^{1}$ & 28.9 & 6.3 & 1.184 & $2.34\cdot 10^{1}$ & 30.3 & 6.5 & 1.184 & $2.17\cdot 10^{1}$ & 34.4 & 9.6 & 1.181\\
0.60 & 0.70 & $1.19\cdot 10^{1}$ & 33.1 & 5.6 & 1.186 & $1.06\cdot 10^{1}$ & 36.4 & 10.0 & 1.188 & $9.13\cdot 10^{0}$ & 36.0 & 10.9 & 1.190\\
0.70 & 1.00 & $2.06\cdot 10^{1}$ & 37.0 & 5.1 & 1.185 & $2.12\cdot 10^{1}$ & 40.4 & 8.8 & 1.182 & $1.84\cdot 10^{1}$ & 39.0 & 6.7 & 1.178\\
\bottomrule
\end{tabular}
\caption{Same as Table \ref{tab:GIM2D_Q1} for \tb cross sections in the range $1122<\Qsq<20000$~\GeVsq. The binning follows the reduced \tb binning given in Table~\ref{tab:binnings}.                  
}
\label{tab:GrTau2D_Q5}
\end{table*}

\end{document}